\documentclass[journal]{IEEEtran}
\usepackage{palatino,epsfig,latexsym,natbib}

\usepackage[T1]{fontenc}
\usepackage{graphicx}
\usepackage{algorithm}
\usepackage{algpseudocode}
\usepackage{amsmath,amssymb,amsfonts}
\usepackage{multirow}

\usepackage[caption=false, font=normalsize, labelfont=sf, textfont=sf]{subfig}

\begin{document}

\title{Multi-objective Evolutionary Approach for Efficient Kernel Size and Shape for CNN }  

\author{Ziwei~Wang,
	Martin~A.~Trefzer,
	Simon~J.~Bale,
	and~Andy~M.~Tyrrell,
	\thanks{The authors are with the Intelligent Systems and Nanoscience Research Group, Department of Electronic Engineering, University of York, York, YO10 5DD, United Kingdom. e-mail: (zw960, martin.trefzer, simon.bale, andy.tyrrell)@york.ac.uk}
}

\maketitle

\begin{abstract}

While state-of-the-art development in Convolutional  Neural  Networks (CNNs) topology, such as VGGNet and ResNet, have become increasingly accurate, these networks are computationally expensive involving billions of arithmetic operations and parameters. 
In order to improve the classification accuracy, state-of-the-art CNNs usually involve large and complex convolutional layers. However, for certain applications, e.g. Internet of Things (IoT), where such CNNs are to be implemented on resource-constrained platforms, the CNN architectures have to be small and efficient. To deal with this problem, reducing the resource consumption in convolutional layers has become one of the most significant solutions. 
In this work, a multi-objective optimisation approach is proposed to trade-off between the amount of computation and network accuracy by using Multi-Objective Evolutionary Algorithms (MOEAs). The number of convolution kernels and the size of these kernels are proportional to computational resource consumption of CNNs. Therefore, this paper considers optimising the computational resource consumption by reducing the size and number of kernels in convolutional layers. Additionally, the use of unconventional kernel shapes has been investigated and results show these clearly outperform the commonly used square convolution kernels. The main contributions of this paper are therefore a methodology to significantly reduce computational cost of CNNs, based on unconventional kernel shapes, and provide different trade-offs for specific use cases. The experimental results further demonstrate that the proposed method achieves large improvements in resource consumption with no significant reduction in network performance. Compared with the benchmark CNN, the best trade-off architecture  shows a reduction in multiplications of up to 6X and with slight increase in classification accuracy on CIFAR-10 dataset.

\end{abstract}

\begin{IEEEkeywords}

Convolutional Neural Networks, 
Evolutionary Computation, 
Deep Learning,
Multi-Objective Evolutionary Optimisation

\end{IEEEkeywords}

\section{Introduction}\label{intro}

Deep Neural Networks (DNNs) are biologically-inspired computing systems which consist of internal connections between artificial neurons. In recent years, DNNs have drawn increasing attention in various application domains, such as image processing, speech recognition and many other challenging Artificial Intelligence (AI) tasks~\citep{guo2016deep}. In particular, Convolutional Neural Networks (CNNs) have successfully gained outstanding performance in image classification and video recognition~\citep{lecun1998gradient,zhao2019object, ren2017faster}. In general, the common CNN architecture usually consists of convolutional layers and pooling layers in between its input layer and fully-connected output layer. The convolutional layers and pooling layers in CNNs are sparsely connected. Compared with the conventional fully-connected layer in a feed-forward neural network, neurons in a convolutional layer are only connected with several neurons in the previous layer. Specifically, any neuron in the feature map of a convolutional layer is a linear combination of neurons in the receptive field which is defined by a convolution kernel in the previous layer~\citep{goodfellow2016deep, yu2015multi}. The state-of-the-art achievements of designing CNN architectures have demonstrated significant improvements on network classification accuracy on various benchmark datasets, such as GoogleNet~\citep{szegedy2015going} and AlexNet~\citep{krizhevsky2012imagenet}. In order to get the best accuracy on specific tasks, the computational complexity of the network grows exponentially and this brings new challenges of executing it on resource-constrained platforms. Based on the analysis by \cite{canziani2016analysis}, several low-cost network architectures can still achieve reasonable accuracy on state-of-the-art benchmark datasets. The motivation to find CNN architectures that can achieve high accuracy and keep computational resource usage at a minimum is therefore high.  

Multi-objective evolutionary algorithms (MOEAs) are today one of the most commonly used methodologies in solving hard optimisation problems where two or more (multiple) conflicting objectives must be satisfied simultaneously~\citep{coello2001short}. MOEAs aim to generate a set of possible solutions (so-called population) in a single run of the algorithm. In recent years, there have been several research studies focusing on optimising neural network typologies and their connection weights. These approaches have demonstrated that using evolutionary algorithms to evaluate neural network topology can achieve competitive performance on state-of-the-art benchmark datasets~\citep{xie2017genetic,suganuma2017genetic,moriarty1998hierarchical}. 

In this article, we investigate optimisation of feed-forward CNN architectures for use in resource-constrained scenarios using MOEAs. In order to achieve a high classification accuracy, state-of-the-art CNN architectures are extremely complex. For example, AlexNet~\citep{krizhevsky2012imagenet} requires billions of multiply-accumulation (MAC) operations to process a single image. In general, convolutional layers are the most computationally expensive ones, where the resource consumption in each layer is proportional to the number and sizes of the convolution kernels. For instance, a \(5\times5\) convolution kernel requires more than twice the number of MAC processes than a \(3\times3\) kernel. Therefore, to reduce the resource consumption in convolutional layers, we aim to reduce the number of MAC processes in these layers by using unconventional kernel shapes as well as reducing the number of kernels in each convolutional layer with minimum loss of network accuracy. The conventional approach to designing the convolutional layer in CNNs is using a set of square convolution kernels which extract the image features. However, the conventional kernels are computationally costly where a large number of multiplications is needed to calculate each feature map. \cite{sironi2014learning}explore the idea of separable convolutions, where a set of smaller 1D convolution kernels is designed to replace the conventional 2D ones. For example, a specific \(n\times n\) matrix can be replaced by the product of a \(n\times 1\) matrix times a \(1\times n\) matrix, instead of calculating one convolution with 9 MAC processes, the reformulation calculates two convolutions with 3 MAC processes each, totally 6 MAC processes, to achieve the same performance for specific tasks. The caveat is that this separation only works within certain constraints, i.e. only a subset of matrices with specific properties can be replaced. Our proposed design contains a number of differently shaped kernels, including 1-D kernels, 2-D rectangle kernels and 2-D square kernels. Each of these kernels can have a different shape and size which require different numbers of MAC processes to extract the features from the input images. In order to minimise the resource consumption and retain the network accuracy while processing the CNNs on the target hardware platform, we use the fast Non-dominated Sorting Genetic Algorithm (NSGA-II)~\citep{deb2002fast} to explore the design space for the feed-forward CNN architecture and produce the best trade-off between network complexity and classification accuracy. Our proposed method is focusing on optimising the network width using combinations of various unconventional shapes and sizes of kernels, while keeping the depth constant.
Moreover, to further reduce resource consumption of convolutional layers, the optimisation approach described in this paper also aims to minimise the number of kernels used in the convolutional layers. The proposed method is tested on widely-used benchmarks, the MNIST~\citep{lecun2010mnist}, Fashion-MNIST~\citep{xiao2017fashion} and CIFAR-10~\citep{krizhevsky2010convolutional} datasets, and compared with conventional CNNs. The proposed method produces a range of network architectures, with some solutions achieving a factor of 5.93X saving in computational resource consumption with a 0.09\% improvement on classification accuracy on CIFAR-10 benchmark dataset. This paper proposed a methodology that can significantly reduce computational cost of CNNs, based on unconventional kernel shapes, and provide different trade-offs for specific use cases. 

The paper is organised as follows: Section~\ref{related_works} gives an overview of the current approaches of optimising feed-forward neural network architecture, as well as optimising learning algorithms and hyperparameters. Section~\ref{design} describes the design methodology that implements NSGA-II to explore the design space of CNN architecture and explains its features. Section~\ref{experiments} shows several test experiments with varying benchmark datasets and provides a proof-of-concept of the performance of the proposed method. Finally, Section~\ref{conclusion} concludes the paper and discusses further work.

\section{Related Works}\label{related_works}

This section introduces a review of the current approaches of automatic optimisation techniques that are used to find good solutions by network topology optimisation, such as connection weights, network structure.

\subsection{Optimisation of Network Architecture}

Evolutionary Algorithms (EAs) are widely used in optimisation problems with complex fitness landscapes. The main idea of optimising artificial neural networks using EAs is to evolve the synaptic weights and connections of the network~\citep{moriarty1998hierarchical}. NEAT~\citep{stanley2002efficient} is a method that uses genetic algorithms (GAs) to change both connection weights and network structure. Their proposed method encodes each neuron and synaptic weight in the genotype. For each iteration, the GA can either add additional neurons to the network or adjust the input/output connections of a neuron. This method allows the GA to find out the best network topology for the target task. A hypercube-based NeuroEvolution of Augmenting Typologies (HyperNEAT) method has shown advantages when optimising weights for CNNs~\citep{stanley2009hypercube,verbancsics2015image}. However, due to the lager search space of CNN topologies, this method requires huge amounts of computational resources. Therefore, this method is difficult to scale up to state-of-the-art deep neural network architectures, because of their size. \cite{morse2016simple} compared evolutionary algorithms with the stochastic gradient descent (SGD) method for weight optimisation of ANNs. Their results demonstrate that using an evolutionary algorithm to optimise weights achieves competitive results, compared with the traditional SGD method. 

From these previous approaches, it is reasonable to assume that GAs are a suitable method for optimising network topology. However, for network weight optimisation, GAs show almost the same performance as SGD through back-propagation. Therefore, in order to search for efficient neural network architectures and updating network weights at the same time, a combination of GAs and SGD methods have been investigated in recent years. \cite{real2017large} apply GAs to CNN design, where the model is trained by SGD through back-propagation and the architecture is optimised by simple GA. They initialise the starting point as a small model which only consists of a single pooling layer. With each evolutionary step, the model is grown by adding more convolution layers. Their approach is only focusing on network accuracy, therefore, the result shows that as the network accuracy is increased, the computational effort required is increased dramatically. CoDeepNEAT~\citep{miikkulainen2019evolving} is a further extension of NEAT~\citep{stanley2002efficient} where the population is separated into two sub-sets: module and blueprint. The module chromosome is a graph that represents a small ANN and the blueprint chromosome is a graph where each mode contains a pointer to a particular module species. During the evolution, the two sub-sets are combined together to build a larger network, where each mode in the blueprint is replaced with a module chosen randomly from the species to which that mode points. Their results show that the network designed by CoDeepNEAT can achieve competitive accuracy in image classification problems with faster training speed. \cite{kim2017nemo} use a multi-objective evolutionary algorithm (MOEA) to trade-off between classification accuracy and run-time. They adopt NSGA-II to explore the Pareto front of the design space. The network architecture is decoded into two categories: number of outputs in each layer and the total number of convolution layers. Their experimental results show that multi-objective optimisation can further reduce the run-time and achieve better accuracy compared with human-expert design. \cite{xie2017genetic} present a GA solution for searching large-scale CNNs. In their work, the GA is applied to designing the network structure, where the connectivity of each layer is encoded by a binary string representation. This method can be easily modified for different network architectures and includes different types of layers and connectivity. 

Apart from using GAs to optimise the CNN architecture by pre-processing the initial populations, such as pre-defining the layer functionalities and connectivity, there are also some GA-based fully automatic architecture design methods addressed in recent years, e.g. \cite{sun2018automatically}. Their design methodology contains a building block that directly using a skip layer to replace the convolutional layer. The skip layer contains two convolutional layers and one skip connection, where the skip connection connects the input of the first convolutional layer to the output of the second convolutional layer. Then, a GA is applied to searching suitable connections of skip layers and pooling layers. Finally, fully-connected layers are added to the tail of the CNN. Similarly, \cite{suganuma2017genetic} proposes using Cartesian Genetic Programming (CGP)~\citep{miller2008cartesian} to represent deep neural network architectures and to use highly functional modules as the node functions to reduce the search space. There are six different node functions in their design, including convolutional blocks, residual blocks, max and average pooling, etc. The CGP encodes CNNs as directed acyclic graphs with a two-dimensional grid of nodes. Their results demonstrate that the architectures built by CGP outperform most of the hand-designed modules and provide a good trade-off between classification accuracy and the number of parameters. 

\subsection{Optimisation in Training}

A large number of hyperparameters need to be configured when designing and training a neural network. These hyperparameters, such as the learning rate, regularisation coefficients and number of training iterations, have significant impact on the convergence efficiency of the training process when considering computational cost. Grid search~\citep{larochelle2007empirical} and random search~\citep{bergstra2012random} have been previously used to find local optima for the values of hyperparameters in the multi-dimensional search space. \cite{bergstra2012random} have demonstrated an optimisation method for hyperparameters based on random search. In their approach, multiple hyperparameters have been considered for optimising a 3-layer network. They have compared the performance between grid search~\citep{larochelle2007empirical} and random search for optimising neural networks. The results show that optimisation through random search can converge to local optima more efficiently than grid search when given the same computing budget, especially for high-dimensional spaces. 

A recent state-of-the-art optimisation technique for hyperparameters in neural networks is Bayesian optimisation. Different from grid search and random search, where each evaluation is processed by purely random sampling of points in the search space, Bayesian optimisation with Gaussian Process~\citep{rasmussen2003gaussian} uses prior solutions of the objective function to determine which objective in the search space needs to be evaluated next. In this method it is assumed that in each dimension of the search space, the distribution of the objective function can be modelled as a Gaussian distribution centred on the mean value of the objective function separately for each dimension. \cite{snoek2012practical} proposed an optimiser based on Bayesian optimisation which considers nine hyperparameters in a CNN, including learning rate, number of epochs etc. Their approach shows that Bayesian optimisation of hyperparameters in CNNs can outperform human-expert designs in terms of classification accuracy. Additionally, an optimisation method for artificial neural networks with a Gaussian Process search algorithm demonstrates significant reduction in both computing time and error rate, compared with random search and grid search~\citep{dernoncourt2016optimizing}. \cite{swersky2013multi} proposed a multi-task Bayesian optimisation method where the knowledge is transferred between a finite number of correlated tasks. Their method separates a large dataset into several smaller subsets, then uses these subsets to find the best settings of hyperparameters for the full dataset by evaluating the configuration of subsets. 

\subsection{Unconventional Convolutions}

Traditional CNN designs use square kernels to detect the image features. This design method brings significant challenges for computational systems, because the number of arithmetic operations increases exponentially as the network size increases. In order to reduce computational resource usage and speed up large CNNs, recent research using unconventional kernel shapes has focused on approximating existing square-kernel convolutional layers for network compression and acceleration. Recent approaches~\citep{sironi2014learning,jaderberg2014speeding,denton2014exploiting} have demonstrated that some of the 2-D square convolution kernels can be factorised into two 1-D kernels. For example, if a convolutional layer contains a set of~\(n\times n\) 2-D kernels, where~\(n\) represents the kernel height and width, it can be factorised as a sequence of two layers with~\(n\times 1\) and~\(1\times n\) kernels, which uses less computations. Therefore, the 1-D convolution approximation can significantly accelerate the classification speed as well as reducing the number of network parameters. Similarly, \cite{jin2014flattened} introduce this into the training phase by factorising a conventional 3-D convolution kernel into three consecutive 1-D kernels. Their results show that by factorising the 3-D convolution layers, the network can be accelerated by approximately a factor of two while sustaining similar or better classification accuracy than using conventional 3-D convolution kernels.

Another approach to design the convolutional layer is to use multiple sizes of kernels in one layer. GoogleNet~\citep{szegedy2015going} is one of the most accurate CNNs on state-of-the-art benchmark datasets. In order to increase the network accuracy, their work introduces an inception module to increase the network depth and width. Firstly, the inception module contains multiple differently-sized convolution kernels, as well as max pooling operations. Different types of kernels are computing in parallel and extract features at multiple scales. After that, feature maps from different kernels are concatenated to form the input of the next module. In order to reduce the computational effort, a~\(1\times 1\) kernel convolution is inserted into the inception module for dimension reduction. Hence, the network can grow deeper and wider with reasonable increase in computational resource usage. \cite{szegedy2016rethinking} modify the inception module, where the~\(7\times 7\) convolution kernels are replaced by
a sequence of~\(7\times1\)  and~\(1\times 7\) kernels, so that the overall computing resources are reduced when compared with the original design. \cite{ding2019acnet} propose Asymmetric Convolution Blocks (ACBs) to replace the conventional convolution kernels. Typically, ACBs replace the conventional convolutional layer with three parallel layers, where these layers contain~\(n\times n\), ~\(n\times 1\) and ~\(1\times n\) kernels respectively. Finally, the outputs from each layer are summed up to enrich the feature space.

\subsection{Other Methods for Reducing Computational Complexity}

Vector Quantization (VQ) is a method for compressing densely connected layers to make CNNs smaller. VQ is a method that quantises groups of numbers together~\citep{gersho2012vector}. Compared with scalar quantisation, VQ can significantly reduce computational and memory resource usage. This is because scale quantisation needs to address every input data one by one~\citep{sullivan1996efficient}, VQ can process a set of numbers concurrently. The disadvantage is that in order to reduce the computational and memory resource usage, VQ always has some loss of accuracy. \cite{denil2013predicting} introduce that there are redundancies in neural network parameters. Then, they apply VQ to reduce the number of dynamic parameters in deep models by representing the weight matrix as a low-rank product of two smaller matrices. \cite{su2018redundancy} follows the research on reducing network redundancy for implementing artificial neural networks on FPGAs. Their work suggests that the network redundancy can be reduced at the data level as well as the model level. The hardware system, after reducing network redundancies, shows that both logic resources and on-chip memory usage can be reduced. Since deep CNN models involve millions of parameters, \cite{gong2014compressing} have investigated information theoretical vector quantization methods for compressing the parameters of CNNs. Their results show that the VQ method has a significant improvement over existing matrix factorisation methods. After 16 to 24 times compression of state-of-the-art CNNs, the classification accuracy only has a 1\% loss.

There is an alternative approach to reducing both the number of operations and parameters, i.e. pruning. The pruning method aims to reduce the number of parameters and operations in CNNs by permanently dropping less important inner-connections~\citep{hanson1989comparing}. The pruning method allows large predecessor networks to inherit knowledge from a smaller network and maintain comparable accuracy. \cite{han2015deep} introduce a method that prunes the network by thresholding the magnitude of weights. If the absolute magnitude of any weight parameter is less than a scalar threshold value, the weight will be pruned. The weight magnitude-based method shows that with network pruning, it can significantly reduce the memory requirement for storage of the neural network as well as the number of neuron connections, without affecting accuracy. \cite{guo2016dynamic} continued this research applied to magnitude-based pruning methods. The method proposed allows restoration of the pruned weights by iteration.  allows the network to be pruned at each iteration and  can significantly reduce the network complexity for large CNN models without any loss in accuracy.

In summary, although, the overall computational resources are reduced by applying the 1-D convolution kernels, those 1-D convolution kernels also lower the networks' classification accuracy. There are still a gap needed to be filled. Is there a combination of non-square convolution kernels for CNNs that can reduce the computational costs  and remain the same or even higher the classification accuracy at the same time? Furthermore, previous researches only focus on developing 1-D convolution kernels, weather other shapes of unconventional convolution kernels can gain CNN's performance also needs to be investigated. The rest of this paper proposes a new method to address the problem that provide different trade-offs between the CNNs' computational costs and classification accuracy based on finding out the combination of unconventional convolution kernels.

\section{Methodology}\label{design}

The main idea in this work for optimisation of the CNN architecture is to consider how the classification accuracy can be improved (or kept the same) while reducing the computational resources required. Literature of previous investigations suggests that using different shapes of convolution kernels can improve the network performance by detecting multiple-scale features from input images~\citep{ding2019acnet}. In this section, an analysis into how the size and shape of the convolution kernels are processed regarding to their computational resource consumption is conducted. Then, a set of different shapes of convolution kernels is designed to be used in convolutional layers. The MOEA used is NSGA-II~\citep{deb2002fast} to automatically discover efficient network architecture by finding the trade-off between the computational complexity and module classification accuracy on the three benchmark datasets. 

\subsection{Computational Resource Consumption}

For state-of-the-art CNN architectures, the computationally most expensive parts are the convolutional layers~\citep{canziani2016analysis}. In a feed-forward CNN, the convolutional layers are used to extract features from input images. The conventional 2-D convolution process is illustrated in Fig.~\ref{fig_1}. Mathematically, the formulae for calculating a single feature map by convolution operation can be described as~\eqref{conv},
\begin{equation}
    O_{:,:,oc} =\sum_{i=1}^{IC}I_{:,:,i} \otimes K_{:,:,oc} \label{conv}
\end{equation}
where \(O_{:,:,oc}\) is an output feature map of the convolutional layer in the \(oc\)-th channel, I is the input image of the convolutional layer with \(IC\) channels, \(K\) is the set of 2-D convolution kernels in current layer and \(\otimes\) represents the convolution operation. It can be seen from the equation, the calculation consists of repeatedly accumulating multiple kernels to form a number of feature maps, which represent the different characteristics of the input image.

\begin{figure}[ht!]
\centering
\subfloat[\label{1a}]{
\includegraphics[width=0.4\linewidth]{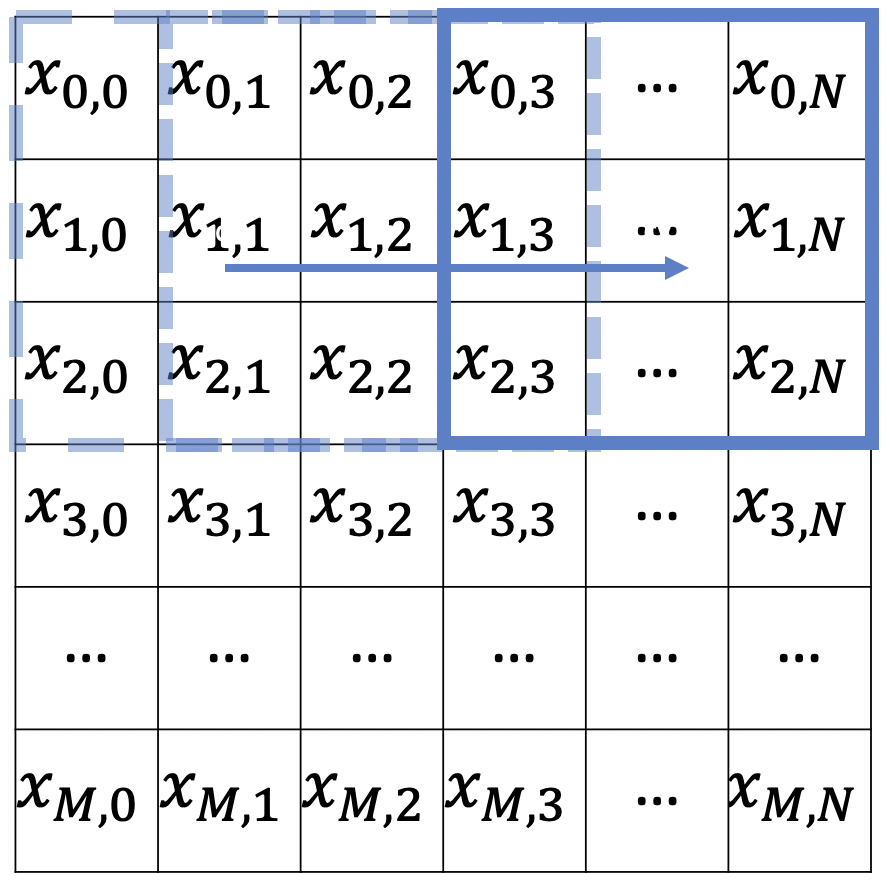}
}
 \hfill
\subfloat[\label{1b}]{
\includegraphics[width=0.4\linewidth]{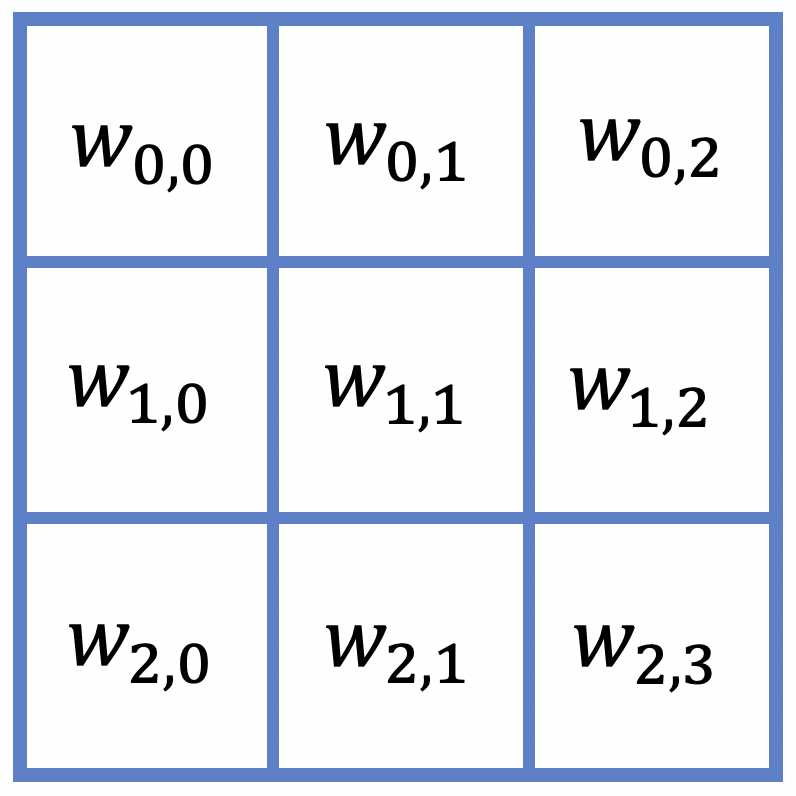}
}
\\

\subfloat[\label{1c}]{
\includegraphics[width=0.8\linewidth]{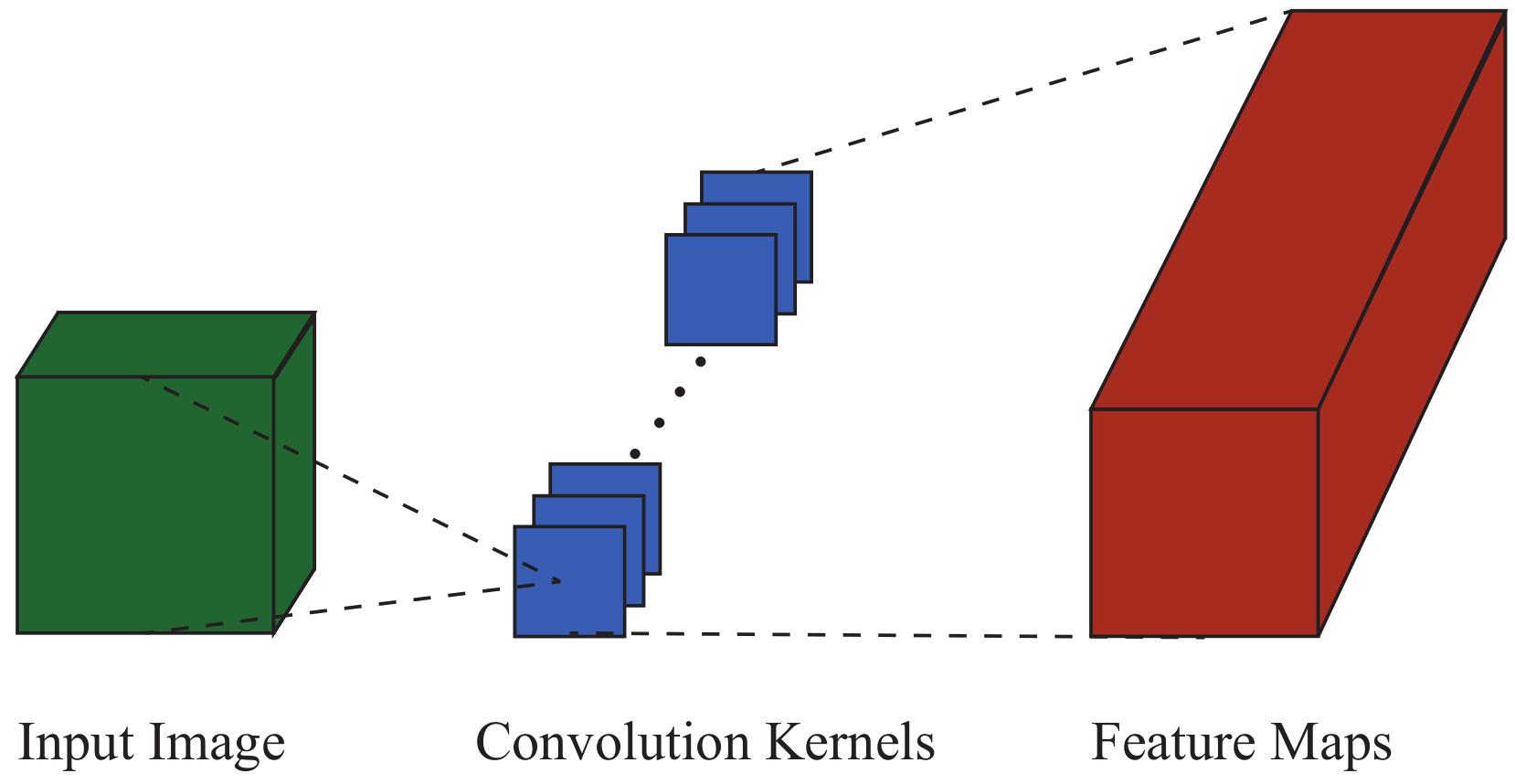}
}
\caption{(a) An example of sliding a $3\times3$ convolutional kernel across the input image. (b) A conventional $3\times3$ square kernel that is used to extract features from the input image. (c) Example of a conventional convolutional layer with multiple square kernels.}
\label{fig_1}
\end{figure}

Processing a CNN in hardware requires multiply-accumulation (MAC) operations to obtain the output feature maps. In order to improve the classification accuracy of CNNs, state-of-the-art CNN architectures have become increasingly complex, therefore, it is necessary for CNN designs to consider their computational resource consumption.

For a convolutional layer, the number of MAC processes is contingent on the size of the convolution kernels, number of kernels and the size of the output feature maps. It can be calculated using following equation:
\begin{equation}
    Operation_{conv} = O_h \times O_w \times O_c \times K_h \times K_w \times K_c\label{conv_operation}
\end{equation}
where \(O_h\) and \(O_w\) indicate the height and width of the output feature maps and \(O_c\) represents the number of output feature maps, i.e. output channels. Similarly, \(K_h\), \(K_w\) and \(K_c\) indicate the height, width and the number of the convolution kernels in the corresponding layer. In the conventional design methodology of CNNs, feature maps and convolution kernels are always square, which means the height and width of the convolution kernels and feature map are the same. 

In a CNN, the fully-connected layer takes the output of the previous convolution processes and predicts the best classification that is labelled to describe the image. Equivalent to~\eqref{conv_operation}, the number of operations in the fully-connected layer of a specific CNN can be calculated as:
\begin{equation}
    Operation_{fc} = O_h \times O_w \times O_c \times N_i\label{fc_operation}
\end{equation}
where the \(O_h\), \(O_w\) and \(O_c\) indicate output feature maps from the last convolutional or pooling layer and \(N_i\) is the number of neurons in the fully-connected layer. 

As can be seen from \eqref{conv_operation} and \eqref{fc_operation}, the convolutional layer requires more MAC operations than the fully-connected layer. Therefore, reducing the number of MAC operations can significantly reduce the computational resource consumption overall. This is important to allow state-of-the-art CNNs to be processed on resource-constrained platforms, such as FPGAs and embedded devices. Apart from the convolutional layers and fully-connected layers, the CNN architecture also involves other layers, including average pooling, max pooling or batch normalisation. For instance, average pooling is used to calculate the average number of pixels in the kernel and max pooling is proposed to find the maximum number of input pixels, which are smaller than operations for convolutional layers in real cases.

\subsection{Mixed Unconventional Kernels}

In this work, in order to minimise the hardware costs while maintaining the CNN classification accuracy, the MAC operations in the convolutional layers are minimised by reducing the kernel sizes, i.e. the product of \(K_h\) and \(K_w\) in \eqref{conv_operation}. For example, a 2-D \(3\times3\) convolution kernel can be replaced by a \(3\times1\) kernel followed by \(1\times3\) kernel, which reduces the number of operations from 9 to 6. However, previous research suggests that the replacement is not equivalent as it does not work as well on some of the lower level layers~\citep{szegedy2016rethinking}, and not all possible \(3\times1\) kernels are captured by the decomposition. Hence, such a substitution requires the network to have extra kernels or layers to compensate, which may potentially increase again the computational complexity. Therefore, the first question to investigate here is what kinds of kernels can be replaced by smaller ones in a convolutional layer. In addition, without adding an additional convolution kernels, it is investigated whether the conventional 2-D square kernels can be directly replaced by more generic sizes of \(m\times n\) shape kernels. Finally, the best combination of different shapes and sizes of kernels for each convolutional layer is considered. 

Utilising different kernel sizes allows the network to extract features at multiple scales~\citep{szegedy2015going}. It is becoming more popular for state-of-the-art CNN architectures to use small kernels, such as \(3\times3\) and \(5\times5\). Therefore, the largest kernel selected for the network architecture proposed here has a size of \(5\times 5\). In order to find the best combination of kernel shapes in a convolutional layer, conventional square kernels and 1-D kernels are used as well as other sizes of kernels, such as \(5 \times 3\) and \(1\times 1\). In total there are 9 different sizes of kernels considered here, the largest being \(5\times 5\) and the smallest one \(1\times1\), as shown in Fig.~\ref{fig_2}.

\begin{figure}[ht!]
\centering
\subfloat[\label{2a}]{
\includegraphics[width=0.5\linewidth]{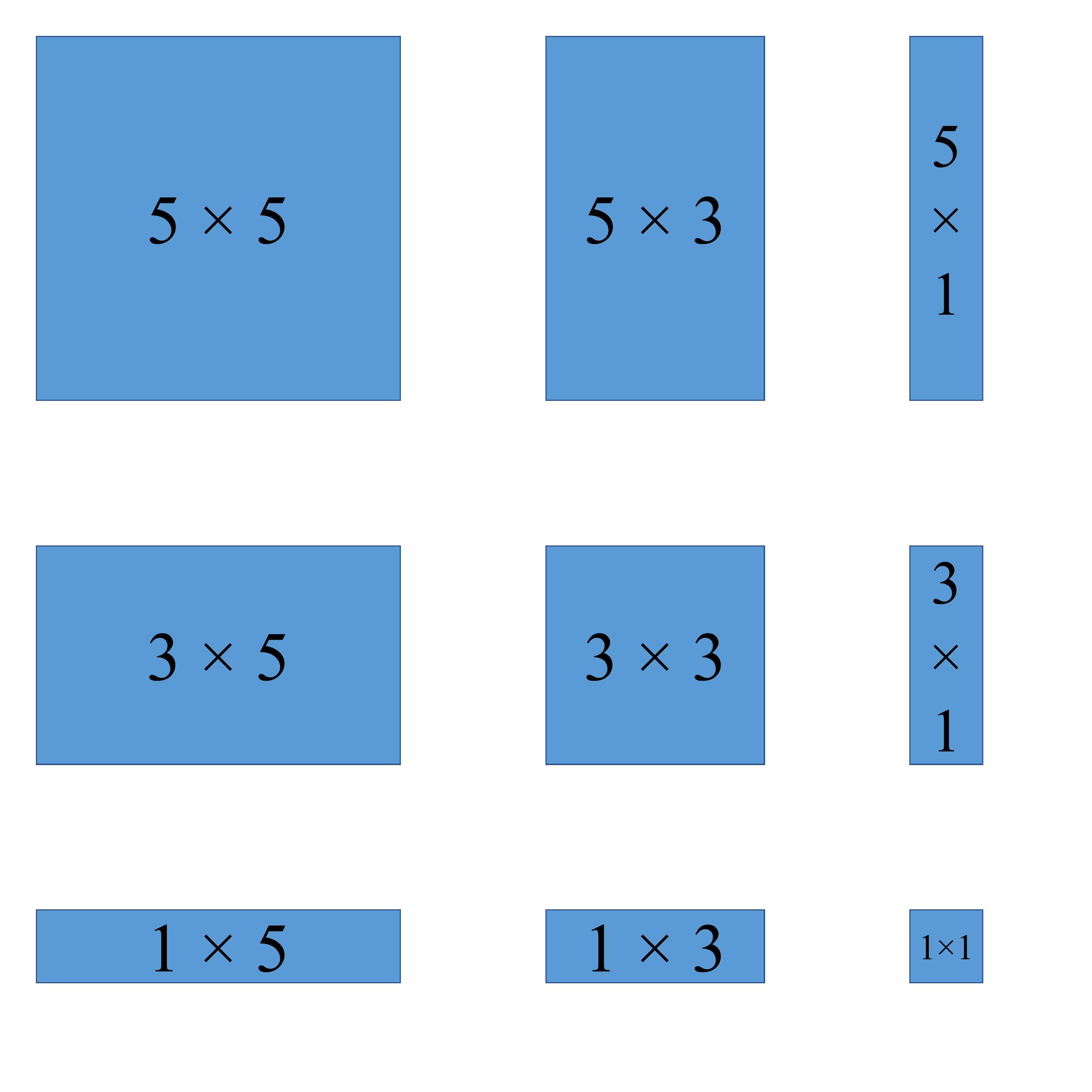}

}

\subfloat[\label{2b}]{

\includegraphics[width=\linewidth]{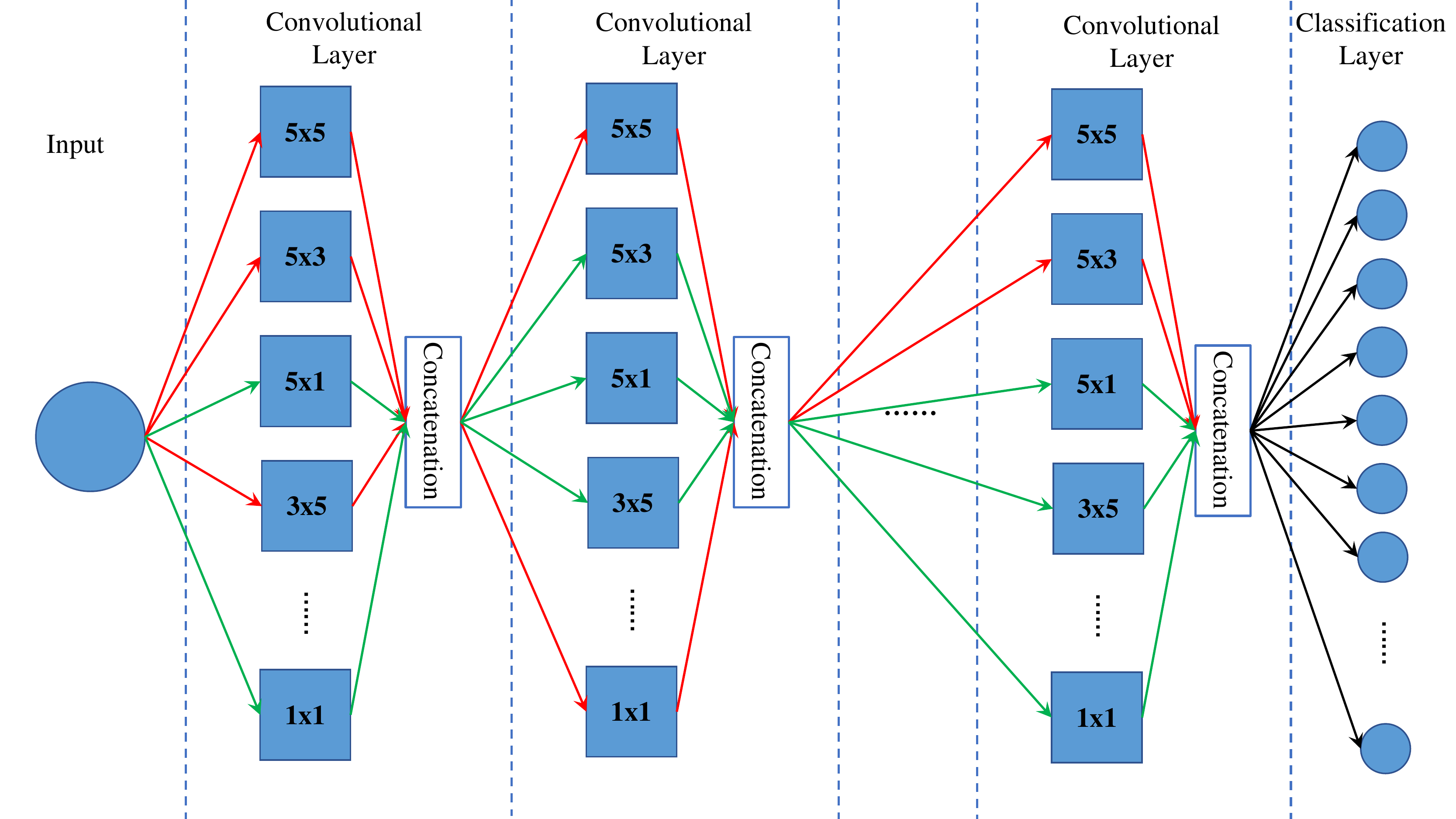}
}
\caption{(a) The set of unconventional kernels. The number of operations required to compute each of the unconventional kernels can be calculated by kernel height $\times$ width. (b) Format of the genotype: The Red and Green Lines shows two different individuals which represent different connections between different size and shape of kernels.  }
\label{fig_2}
\end{figure}

Previous approaches have shown that it is possible to use a series of kernels with differing sizes to better handle multiple-scale objects in a convolutional layer~\citep{szegedy2015going}. Rather than using conventional square kernels, this work puts the set of 1-D stripe kernels into a single convolutional layer. This is so that the network can operate in parallel on different sizes with the most accurate detailing, \(1\times 1\), to the biggest kernels, \(5\times 5\). Then, all feature maps generated from the different kernels are concatenated together into a single output tensor in order to form the input of the next stage. Padding strategy are applied to make sure that the output feature maps from each set of unconventional kernels will have the same resolution as others. Then, the concatenation are used to combine the output feature maps from each set of unconventional kernels into one output tensor. The overall architecture of the proposed design can be viewed as Fig.~\ref{fig_3}.
\begin{figure}[ht!]
\centering

\includegraphics[width=0.5\linewidth]{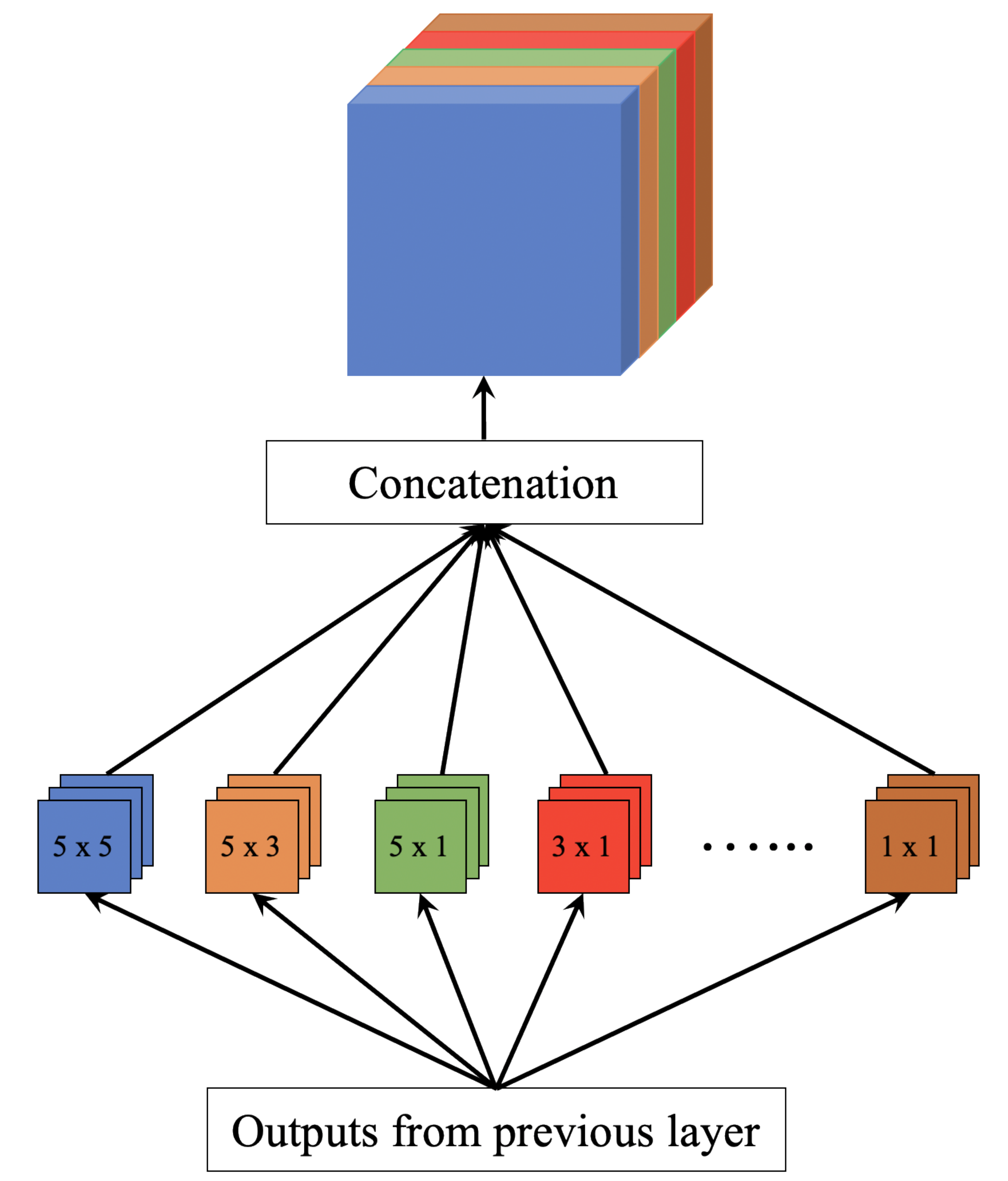}

\caption{Overall design of the convolutional layer which involves multiple sizes of kernels to produce different feature maps. If the strides of each type of kernel are the same, using the padding method, the output from each set of kernels will be the same as other kernels. Therefore, the final output from this layer can be concatenated into a single tensor with no additional computation operation required.}
\label{fig_3}
\end{figure}

\subsection{Multi-Objective Evolutionary Optimisation}

Following the convolutional layer design methodology from the previous section, it is difficult to define the optimal kernel sizes required to replace the conventional square kernels in a given CNN architecture. The new methods described here propose to use the non-dominated sorting genetic algorithm (NSGA-II) to explore the kernel size design space of the CNN architecture. NSGA-II is one of the most popular multi-objective optimisation algorithms which uses a fast non-dominated sorting approach and diversity preservation~\citep{deb2002fast}. The approach of optimising for Pareto optimality makes it possible to trade-off between network accuracy and hardware resource consumption. An overview of optimising a given CNN architecture is shown in Fig.~\ref{fig_4}.

In the optimisation loop, each of the possible unconventional kernel shapes is identified with its kernel ID that represents a specific kernel. As shown in Fig.~\ref{fig_4}, the optimisation loop takes a given conventional CNN architecture as its input. 
\begin{figure}[ht!]
\centering

\includegraphics[width=0.8\linewidth]{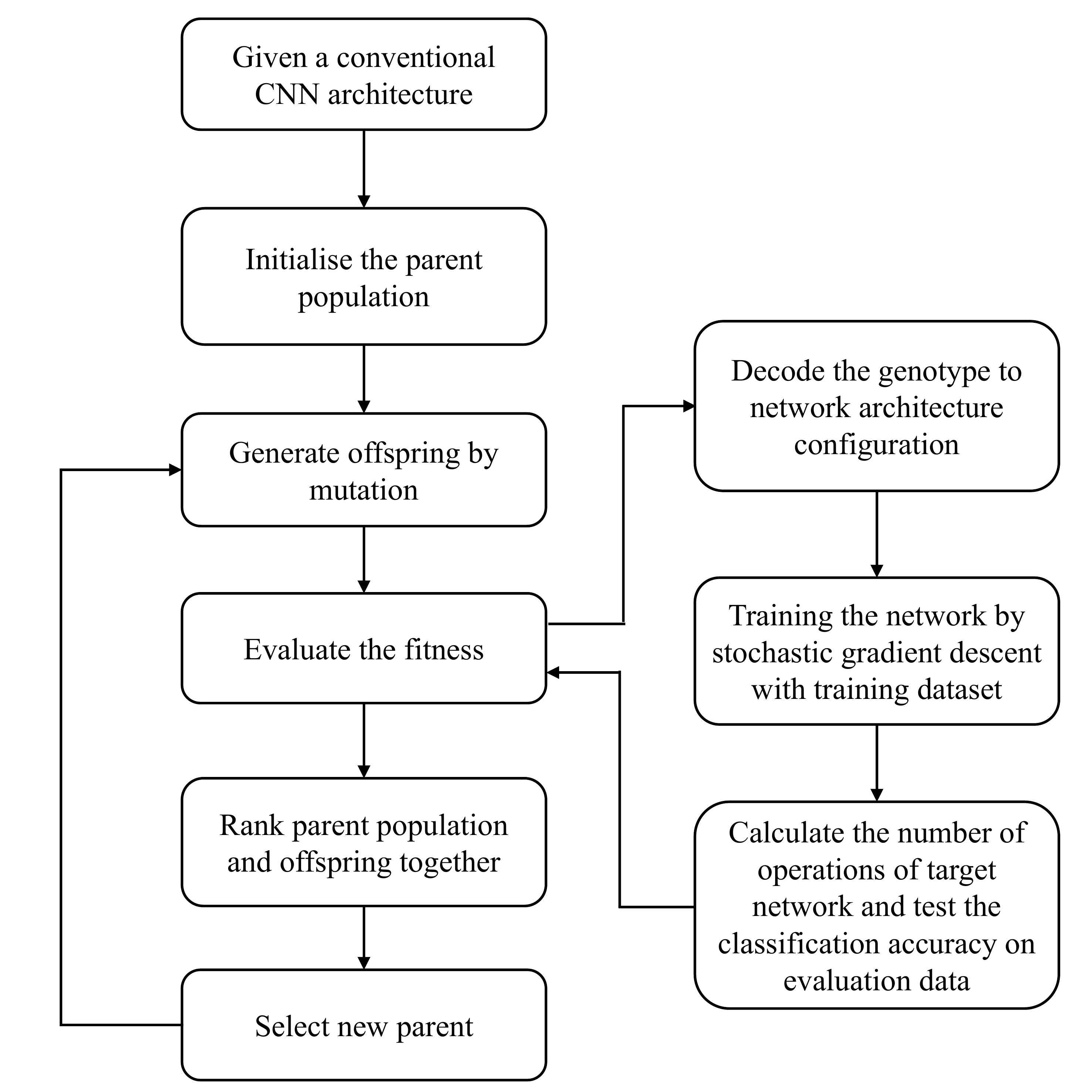}

\caption{Overview of the multi-objective optimisation loop. The method selects a set of unconventional kernel shapes replacing the original (conventionally used) square kernels in a given CNN architecture. Each individual is trained on a training data set. Then, the fitness is calculated and assigned based on the classification accuracy on the evaluation data set (objective 1) and the number of arithmetic operations (objective 2). NSGA-II searches for the Pareto front that trades-off between of number operations and model classification accuracy. }
\label{fig_4}
\end{figure}

The kernels from all layers of the input CNN are encoded in the genotype. As we are focusing on optimising the convolutional layers, other hyperparameters of the network are kept the same as in the input CNN, e.g. other types of layers, stride, activation function and number of layers. The initial parent population of NSGA-II is generated by randomly replacing the original square kernels with randomly selected unconventional kernel sizes for all convolutional layers in the network. After the initial parent population is created, the first offspring population is generated by mutation operation, which changes the shape of randomly selected kernels based on a given probability. In this case, the genetic representation only consists of a single chromosome, a vector encoding the kernel shapes. Crossover operation is not used in this case.

Then, the fitness of each individual in the population is evaluated by calculating the number of operations in the convolutional layers and testing the classification accuracy of the trained model on a validation set. The architecture generated by NSGA-II is trained using stochastic gradient descent (SGD), using a model training dataset. In this design, one of the fitness measures of NSGA-II is the classification accuracy which is defined as the Top-1 accuracy of trained model on the test dataset. Another fitness is the summation of MAC operations required to compute all of the convolutional layers in the network, which is calculated by \eqref{conv_operation}.

Finally, NSGA-II ranks the fitness of each individual by using a non-dominated sorting approach and a diversity preservation strategy to ensure selection with elitism and a uniform spread of solutions. In non-dominated sorting, each solution \(p\) has two entities: the first is domination count, the number of solutions that dominate \(p\); the second is the number of solutions that \(p\) dominates. All solutions will be sorted according to each solution's domination count into multiple ranking levels. Diversity preservation is achieved by adopting a crowding distance comparison. So that, when there are two solutions with the same domination level, the one that resides in less densely populated points of the solution space is selected~\citep{coello2006evolutionary}. Following this, half of the individuals which have higher rank will be selected as the parent population for the next generation.

\section{Experimental Setup and Results}\label{experiments}

Each individual's fitness needs to be evaluated separately by training and testing the resulting network. State-of-the-art CNN architectures may require extremely large computational budgets for processing the networks. The aim in this paper is to show the improvement of the proposed method compared with conventional convolutional layers in terms of trade-off between computation costs and classification accuracy. Therefore, a small CNN, the Lenet-5 architecture, is used here as the benchmark network to illustrate the improvement achieved by our proposed method. To evaluate the capability and the full potential for scalability of the proposed method, it is also tested on deeper CNN architectures. 

\subsection{Experimental Settings}

The benchmark CNN architecture is built based on the Lenet-5 architecture~\citep{lecun1998gradient}. The original Lenet-5 consists of two convolutional layers and two max-pooling layers, a fully-connected layer and a classification layer. In order to improve the classification accuracy of the network, we increase the number of kernels in each of the convolutional layer and nodes in the fully-connected layers. The overall architecture is shown in Fig.~\ref{fig_5}, which is used as the benchmark topology to test the proposed method.

\begin{figure}[ht!]
\centering

\includegraphics[width=0.7\linewidth]{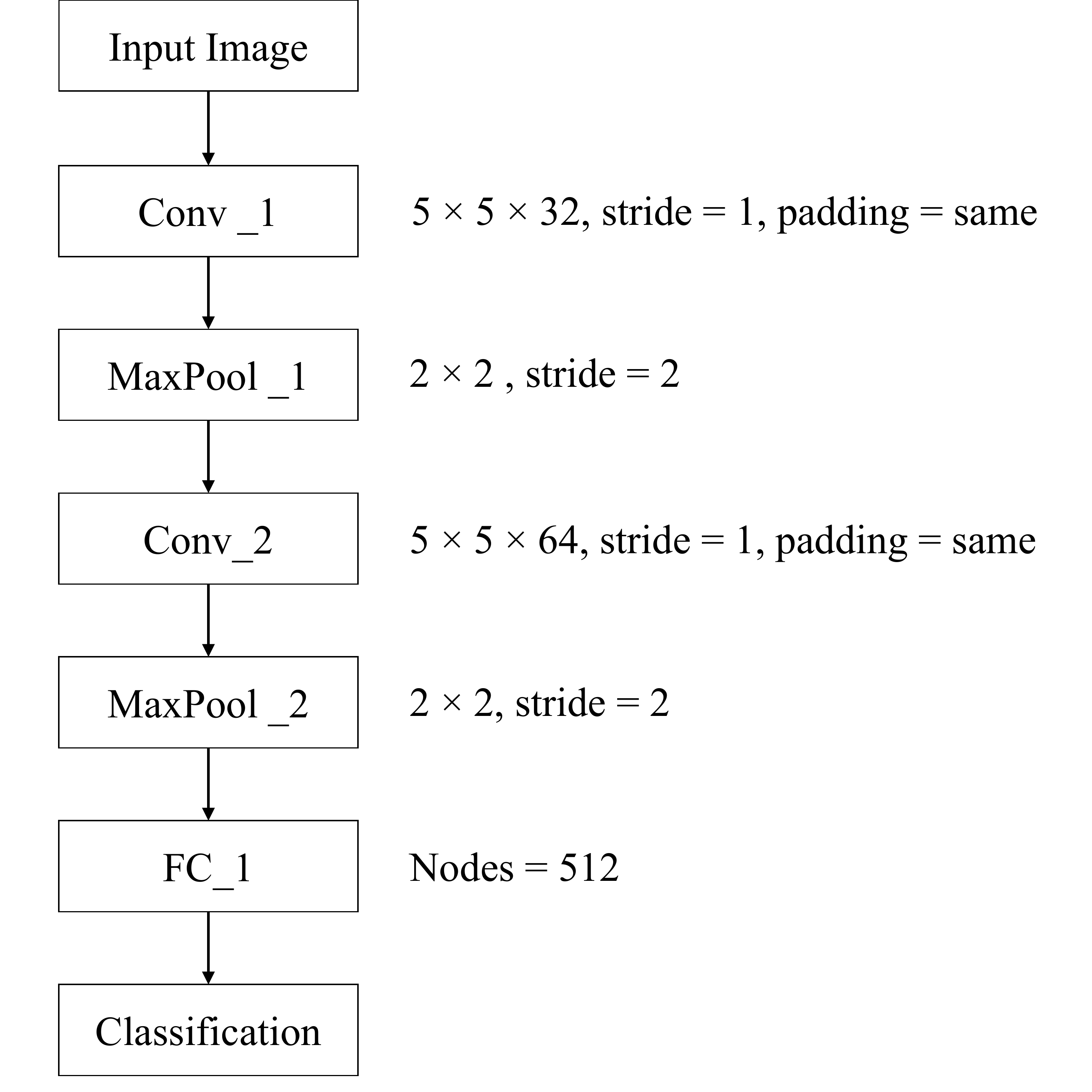}

\caption{The benchmark CNN used to test our optimisation method. The benchmark CNN is based on the Lenet-5 architecture. The network consists two 2-D convolutional layers, which contain 32 and 64 kernels respectively. All of the kernels have dimensions of \(5\times 5\) and the stride is 1. Each convolutional layer is followed by a max pooling one with dimensions of \(2 \times 2\) and a stride of 2. There is a fully connected layer connected to the output of the second max pooling layer that has 512 nodes. Finally, a classification layer is used to predict the best classification label applied to the image. }
\label{fig_5}
\end{figure}

The optimisation method is applied to three different datasets, MNIST, Fashion-MNIST and CIFAR-10. MNIST and Fashion-MNIST consist of a training set of 60,000 images and a test set of 10,000 images. Each image in the two MNIST sets is a \(28\times 28\) grayscale image, associated with a label from 10 different classes. CIFAR-10 is split into a training set of 50,000 images and a test of 10,000 images. Each image in the CIFAR set is a \(32\times 32\) pixel RGB image, associated with a label from 10 different classes. The network is trained on the training sets and evaluated in the test set, which is one of the fitness measures of the proposed method. The number of operations is calculated by the total number of multiplications in two convolutional layers, the second objective measure used here. All of the networks are trained by SGD method and use the Adam optimiser~\citep{kingma2014adam} with a learning rate of 1e-3. The softmax cross-entropy loss is used as the loss function. Each model is trained for 30 epochs.

In order to explore the Pareto front of the CNN architecture, the optimisation loop is set to run 100 generations for a population size of 25 individuals. There is only one type of chromosome. Hence, only mutation is used as the genetic operator, the mutation rate is 0.1. 

\subsection{Two-Objective Optimisation}

The proposed method is evaluated on CIFAR-10 dataset using two optimisation objectives, number of multiplications in convolutional layers and network classification accuracy. In this experiment, the training dataset is created by randomly selecting 40,000 images from the training set, and the remaining 10,000 images are used for fitness evaluation. Finally, architectures found are trained on the training set for 100 epochs and classification accuracy is tested on test set. To prevent overfitting, a weight decay of 0.0001 and data augmentation have been used for training the networks. The data augmentation used is based on \cite{he2016deep}, that is padding 4 pixels on each side and randomly crop a patch from the padded image or its horizontally flipped version. In order to handle the colour image inputs, the input layer of the benchmark network is modified as 3 channel input.

After configuring the benchmark network to accept colour images, the total number of multiplications required for the convolutional layers further increases to 15,564,800, and the classification error is 17.63\% on CIFAR-10. The optimisation loop has a population of 25 individuals and was run for a 100 generations. Three different architectures from the set of solutions are chosen as reference points. The first reference point, Ref 1, has the highest accuracy. The second reference point is one which involves significantly less computational resource while still featuring high accuracy. The third reference point is the trade-off solution closest to the origin between number of multiplications and classification accuracy. The optimisation results and details of kernels usage of three reference architectures can be viewed in Appendix A.

These three architectures are then re-trained on the whole training set, i.e. 50,000 images for training, and the classification accuracy is evaluated on the test set which contains 10,000 examples. Each architecture is re-trained for 100 epochs with weight decay and data augmentation. All of them are trained by using Adam optimiser~\citep{kingma2014adam} with initial learning rate of 0.001 and the learning rate is reduced by factor of 10 at 30th epoch. After re-training and testing, the results are shown on Table~\ref{tab_1}. The comparison between the benchmark network and reference point is shown in Table~\ref{tab_1}.

Then, the proposed method is tested on MNIST and Fashion-MNIST datasets. With the same training settings, the benchmark network has a classification accuracy of 98.92\% on MNIST and 98.92\% on Fashion-MNIST. Both benchmark networks require a total of 10,662,400 multiplications in their convolutional layers. There are three reference solutions that have been selected from each set of solutions using the same approach as before: the first reference solution features the highest accuracy, the second reference solution uses less resources while featuring high accuracy, and the third reference solution has the closest-to-origin trade-off between number of multiplications and classification accuracy. The full optimisation results and details of kernels usage of three reference architectures in each dataset can be viewed in Appendix A.

\begin{table}[htbp!]
\caption{Comparison between the benchmark network and solutions found by the proposed method on CIFAR-10, MNIST and Fashion-MNIST datasets. Three reference points are selected from optimised results for each dataset. The full optimisation results and details of kernels usage of three reference architectures in each dataset can be viewed in Appendix A.}
\begin{center}
\resizebox{\columnwidth}{!}{%
\begin{tabular}{c c c c c}
 \hline
 Dataset          &Model          & Top-1 Acc.  & Acc. improve   & Reduction in Mults.\\
 \hline

 \multirow{4}{*}{MNIST}         &Benchmark      & 98.92\%               & -                           & - \\
                                &Ref 1          & 99.56\%               & 0.64\%                      & 2.15x \\
                                &Ref 2          & 99.54\%               & 0.62\%                      & 3.13x \\
                                &Ref 3          & 99.49\%               & 0.57\%                      & 7.02x \\
 \hline
 \multirow{4}{*}{Fashion-MNIST} &Benchmark      & 92.12\%               & -                           & - \\
                                &Ref 1          & 93.14\%               & 1.02\%                      & 2.95x \\
                                &Ref 2          & 93.07\%               & 0.95\%                      & 4.24x \\
                                &Ref 3          & 92.79\%               & 0.67\%                      & 9.15x \\
 \hline
\multirow{4}{*}{CIFAR-10}      &Benchmark      & 82.37\%               & -                           & - \\
                                &Ref 1          & 83.75\%               & 1.38\%                      & 2.71x \\
                                &Ref 2          & 82.54\%               & 0.17\%                      & 4.06x \\
                                &Ref 3          & 80.84\%               & -1.53\%                     & 7.14x \\
                                
  \hline
\end{tabular}
}
\label{tab_1}
\end{center}
\end{table}

\subsection{Three-Objective Optimisation}

Based on the experimental results reported in the previous section using two objectives, an additional step is considered with the aim to further reduce the number of multiplications in convolutional layers by allowing the proposed method to remove kernels if possible. A third objective has been added to the optimisation loop accordingly, which is the number of kernels. As described in Section~\ref{design}, the number of operations of convolutions will also be reduced by using fewer kernels. Therefore, the number of kernels is a secondary measure that puts optimisation pressure on reducing the resource consumption of the networks. The proposed method is now aiming to find the trade-off between three objectives, i.e. number of multiplications used in convolutional layers, model classification accuracy and number of kernels used in each layer. The benchmark CNN is the same LeNet-5 architecture with the original set of kernels as used in the previous experiments.

\begin{figure}[ht!]
\centering
\subfloat[\label{9a}]{
\includegraphics[width=0.75\linewidth]{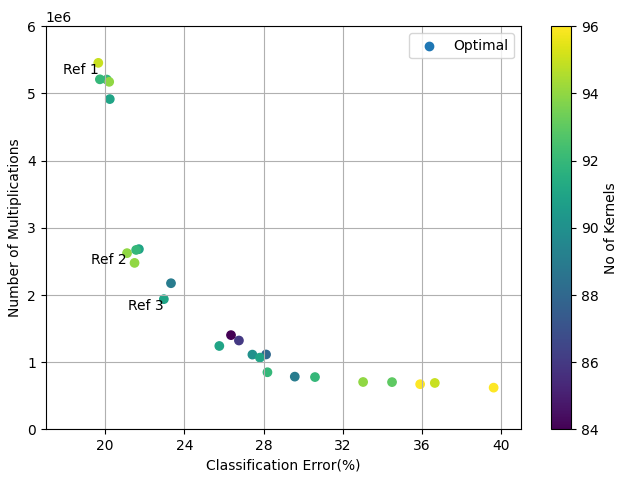}
}
\hfill
\subfloat[\label{9b}]{
\includegraphics[width=0.75\linewidth]{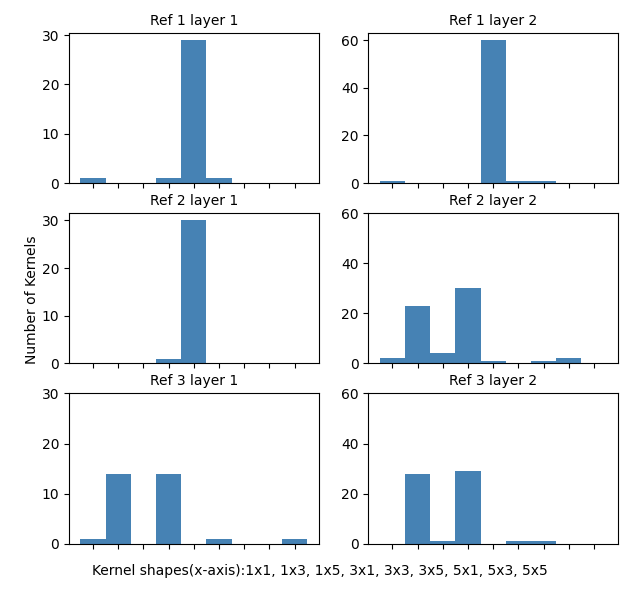}
}

\caption{(a) Optimised architectures by the proposed method after 100 generations on CIFAR-10. (b) The kernel distribution of three reference points on (a). }
\label{fig_9}
\end{figure}

In this case, two different architectures are chosen which are found by the 3-objective method from each dataset as reference points and compared with the benchmark network that has been used in previous comparisons. The results for optimising two layer network with three objective optimisation on MNIST, Fashion-MNIST and CIFAR-10 are shown in Table~\ref{tab_4}.  

\begin{table}[htbp!]
\caption{Comparison between the benchmark network and solutions found by proposed method on three datasets with three-objective optimisation. Three reference points are selected from optimised results for each dataset. The full optimisation results and details of kernels usage of three reference architectures in each dataset can be viewed in Appendix B.}
\begin{center}
\resizebox{\columnwidth}{!}{%
\begin{tabular}{c c c c c}
 \hline
  Dataset          &Model          & Top-1 Acc.  & Acc. improve   & Reduction in Mults.\\
 \hline
\multirow{4}{*}{MNIST}          &Benchmark      & 98.92\%               & -                             & - \\     
                                & Ref 1         & 99.52\%               & 0.60\%                        & 2.84x \\
                                & Ref 2         & 99.48\%               & 0.56\%                        & 7.04x \\
                                & Ref 3         & 99.38\%               & 0.46\%                        & 18.38x\\
 \hline
\multirow{4}{*}{Fashion-MNIST}  &Benchmark      & 92.12\%               & -                             & - \\    
                                & Ref 1         & 92.96\%               & 0.84\%                        & 3.74x \\
                                & Ref 2         & 92.75\%               & 0.63\%                        & 8.12x  \\
                                & Ref 3         & 92.67\%               & 0.55\%                        & 11.03x \\
 \hline
\multirow{4}{*}{CIFAR-10}      &Benchmark       & 82.37\%               & -                             & - \\     
                               & Ref 1          & 83.47\%               & 1.10\%                        & 2.85x \\
                               & Ref 2          & 82.46\%               & 0.09\%                        & 5.93x \\
                               & Ref 3          & 80.73\%               & -1.64\%                       & 8.03x \\
  \hline
\end{tabular}
}
\label{tab_4}
\end{center}
\end{table}

Fig.~\ref{9a} illustrates optimal solutions for the benchmark network on the CIFAR-10 dataset.  Compared with the two-objective optimisation in the previous sections, a number of kernels in both convolutional layers have been removed by the three-objective optimisation method. By removing some of the kernels, optimising for the three objectives can further reduce the number of multiplications that are required to process the CNN without compromising significantly on classification accuracy. In addition, Fig.~\ref{9b} shows the kernel distributions of three reference points for both convolutional layers after applying the proposed method. The first reference point is the network architecture which features the highest classification accuracy achieved by the proposed method. The second reference point is the one that still features good classification performance while uses less computational resources than the first reference point. Notably, it also outperforms the benchmark network architecture in this case. The third reference point is the best trade-off (closest to origin) between number of multiplications, number of kernels and classification accuracy after optimisation. Details of kernels usage of three reference architectures can be viewed in Appendix B.

These networks from the three reference points have been re-trained for 100 epochs on the whole training set of CIFAR-10, i.e. 50,000 images for training, and the classification accuracy is evaluated on the test set which contains 10,000 examples. It can be seen from Fig.~\ref{9b}, Ref 1 contains a majority of \(3\times 3\) kernels, and combined with a small number of unconventional kernels in both convolution layers. For Ref 2 and Ref 3, these networks mainly involve \(3\times 3\) kernels in the first convolutional layer and also make use of kernels with different sizes. In the second convolution layer of Ref 2, most of \(3\times 3\) kernels are replaced by \(1\times 3\) kernels and \(3\times 1\) kernels. This replacement is the main reason for Ref 2's significant reduction in computational resource compared Ref 1 and only has a small impact on its classification accuracy. Both the Ref 1 and Ref 2 outperform the benchmark CNN describe in Fig.~\ref{fig_5}. Ref 3 provide the best trade-off between the number of multiplications, number of kernels and classification accuracy. It can be seen that Ref 3 uses 8.03x fewer multiplications than the benchmark CNN with only 1.64\% increases of classification error. Both convolutional layers in Ref 3 involve \(1\times 3\) kernels and \(3\times 1\) kernels at the most.

The three-objective optimisation has also been tested on MNIST and Fashion-MNIST datasets. The optimisation results of the benchmark CNN on MNIST dataset are shown in Fig.~\ref{8a} and Fig.~\ref{8b} illustrates the optimisation results for the benchmark CNN on MNIST and Fashion-MNIST, respectively. Details of kernels usage of three reference architectures in each dataset can be viewed in Appendix B.

\begin{figure}[ht!]
\centering
\subfloat[\label{8a}]{
\includegraphics[width=0.75\linewidth]{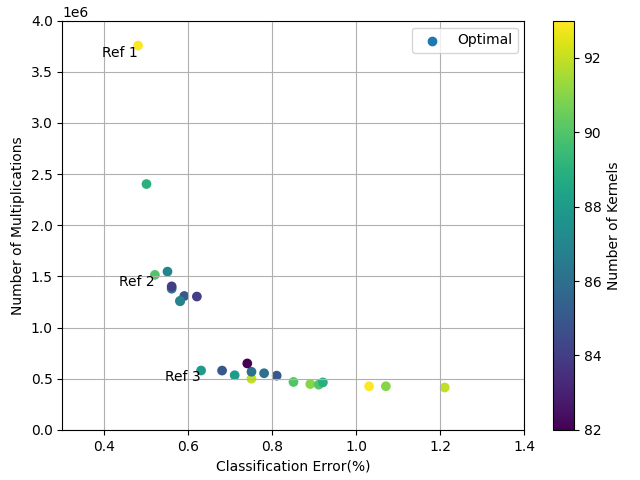}
}
\hfill
\subfloat[\label{8b}]{
\includegraphics[width=0.75\linewidth]{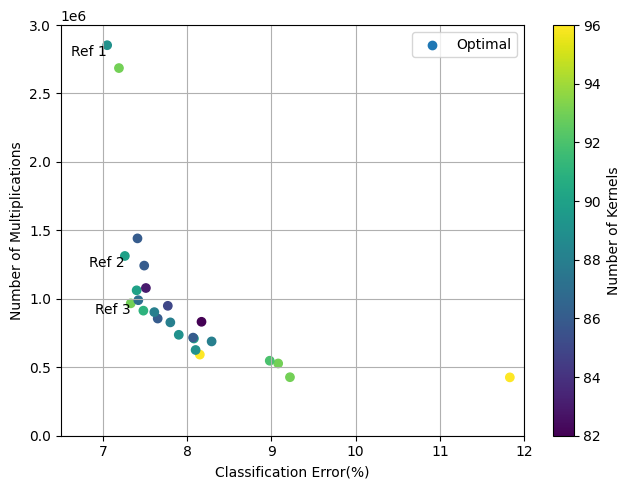}
}

\caption{(a) Optimised architectures by the proposed method after 100 generations on MNIST. (b) Optimised architectures by the proposed method after 100 generations on Fashion-MNIST.. }
\label{fig_8}
\end{figure}

\subsection{Results on Deeper CNNs}

As shown in the results of the experiments, the proposed method demonstrates significant improvements in reducing computational resource consumption in convolutional layers. The effectiveness of the proposed method is now evaluated on deeper CNNs to give some indication of scalability. The benchmark architecture that is used in these experiments are three and four convolutional layer architectures. Based on the previous experiments, the benchmark architectures have been modified to use \(3\times 3\) kernels in each convolutional layer as a starting point. 

The three-layer benchmark architecture consists of three convolutional layers, each layer includes 64 \(3\times 3\) convolution kernels. Each convolutional layer is followed by a \(2 \times 2\) max pooling operation. The four-layer benchmark architecture has four convolutional layers, each layer also includes 64 \(3\times 3\) convolution kernels. The max polling operations are applied at the end of the first, second and fourth convolutional layers. After the convolutional layers, there is a fully-connected layer with a total of 512 neurons and a sofmax operation is used to predict the classification. These experiments use three objectives to evaluate the fitness, which are number of multiplications, Top-1 classification error and number of convolution kernels. During optimsation, all networks are trained on CIFAR-10 dataset for 30 epochs and the evolutionary loop is set to run 100 generations with population size of 25. Three reference architectures from each experiment are selected and re-trained for 100 epochs on the whole training set of CIFAR-10 and the classification accuracy is evaluated on the test set. The first reference point features the highest classification accuracy in the solution set after optimisation. The second reference point involves fewer multiplications and still features good classification accuracy. The third reference point is the best trade-off solution (closest to origin) between number of multiplications, number of kernels, and classification accuracy. Table.~\ref{tab_5} illustrates the results of three reference architectures that are found by the proposed method for each benchmark. Fig.~\ref{fig_10}  and Fig.~\ref{fig_11} show the optimal results that are found by the proposed method on CIFAR-10 dataset. Details of kernels usage of three reference architectures can be viewed in Appendix C.

\begin{table}[htbp!]
\caption{Experimental results for three and four convolutional layers network on CIFAR-10}
\begin{center}
\resizebox{0.9\columnwidth}{!}{%
\begin{tabular}{c c c c}
 \hline
 Model         & Referent Network & Acc. improve   & Reduction in Mults.\\
 \hline
 \multirow{3}{*}{three-layers}      & Ref 1                    & 0.16\%                        & 1.17x \\
                                    & Ref 2                    & -1.03\%                        & 2.31x \\
                                    & Ref 3                    & -2.44\%                        & 3.06x \\
                             \\
 \multirow{3}{*}{four-layers}       & Ref 1                    & -0.32\%                        & 2.13x \\
                                    & Ref 2                    & -0.95\%                        & 2.90x \\
                                    & Ref 3                    & -2.46\%                        & 4.42x \\

  \hline
\end{tabular}
}
\label{tab_5}
\end{center}
\end{table}

\begin{figure}[ht!]
\centering
\subfloat[\label{10a}]{
\includegraphics[width=0.75\linewidth]{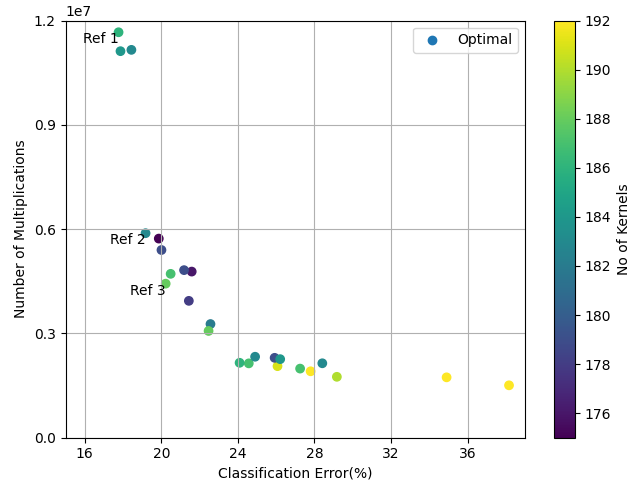}
}
\hfill
\subfloat[\label{10b}]{
\includegraphics[width=0.75\linewidth]{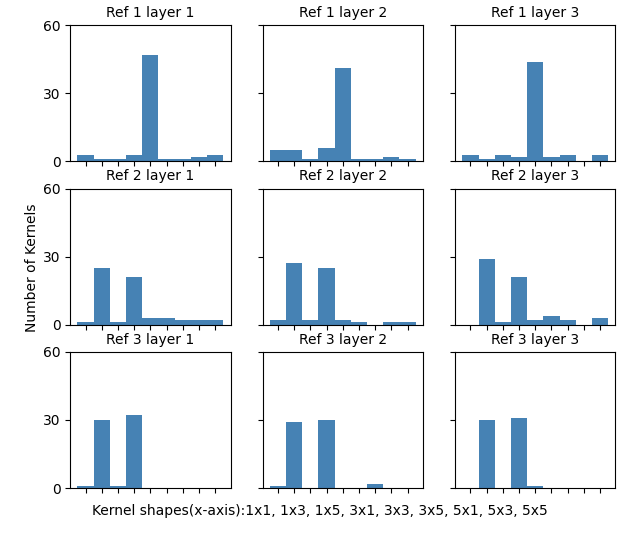}
}
\caption{(a) The three-layer architectures that optimised by the proposed method after 100 generations on CIFAR-10. (b) The kernel distribution of three reference points on (a). }
\label{fig_10}
\end{figure}

\begin{figure}[ht!]
\centering

\includegraphics[width=0.8\linewidth]{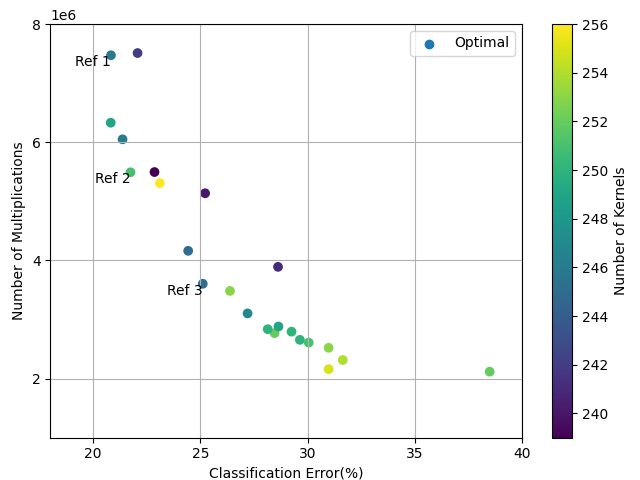}

\caption{The four-layer architectures that optimised by the proposed method after 100 generations on CIFAR-10.}
\label{fig_11}
\end{figure}

As shown in Table.~\ref{tab_5}, the proposed method still can outperform the network using only conventional \(3\times3\) convolutions, which is the standard setting in modern CNN architecture design. As the results show, the number of multiplications in convolutional layers of the three-layer CNN achieves 1.17x less than the benchmark architecture, while the Top-1 accuracy increases 0.16\%. For four-layer CNN, the proposed method achieve 2.13x saving in number of multiplication with 0.32\% decrease in Top-1 accuracy. This demonstrates the capability of the proposed method to be applied to deeper CNN architectures. 

In order to validate the performance of the optimised architecture by the proposed method on various datasets, the best trade-off architectures, Ref 3 from Fig~\ref{10a} and Fig.~\ref{fig_11}, from both three-layer and four-layer networks have been selected to train on MNIST and Fashion-MNIST datasets. These two datasets contain \(28 \times 28\) grey-scale images, so that, benchmark CNN and selected networks are configured as one input channel with \(28 \times 28\) input images. Two networks are trained by using Adam optimiser for 100 epochs with initial learning rate of 0.001 and the learning rate is reduced by a factor of 10 at 30th epoch. The experimental results have been reported in Table.~\ref{tab_6}.

\begin{table}[htbp!]
\caption{Experimental results for three and four convolutional layers network}
\begin{center}
\resizebox{0.8\columnwidth}{!}{%
\begin{tabular}{c c c c}
 \hline
 Model         & Dataset & Original Acc.   & Optimised Acc.\\
 \hline
 \multirow{2}{*}{three-layers}      & MNIST                    & 99.08\%                        & 99.13\% \\
                                    & Fashion-MNIST            & 92.64\%                        & 93.06\% \\
 \hline
 \multirow{2}{*}{four-layers}       & MNIST                    & 99.14\%                        & 99.04\% \\
                                    & Fashion-MNIST            & 92.72\%                        & 92.53\% \\

  \hline
\end{tabular}
}
\label{tab_6}
\end{center}
\end{table}

\section{Conclusion}\label{conclusion}
In this article, we proposed a generic Multi-objective Evolutionary Algorithm (MOEA)-based approach for optimising the size and efficiency of CNN architectures by introducing unconventional (non-square) kernel shapes and combining different sizes of convolution kernels. The proposed method automatically generates combinations of these unconventional kernels that are used to replace the set of one-size square convolution kernels produced by a conventional approach. The optimisation by MOEA provides a trade-off solution space between computational resources and classification accuracy, which is unique to such algorithms. The results show that a significant reduction in the computational resource consumption with negligible sacrifice of (and sometimes slightly increased) performance. 

Moreover, in order to put further emphasis on reduction of the computational resources, the proposed method can reduce the number of kernels used in each convolutional layer in addition to replacing the square kernels by the set of unconventional kernels. 
The proposed method has been tested on MNIST, Fashion-MNIST and CIFAR-10 datasets with CNN architectures of various depth. As can be seen from the results, the proposed method shows large improvements on computational resource consumption, sometimes with increases in classification accuracy, compared with conventional design of convolutional layers. Trading off accuracy with computational complexity of resource consumption in CNNs running in resource-constrained environments is a real-world problem. This significant reduction of computational resources allows deep CNN architectures to be efficiently implemented on many resource-constrained platforms, such as Field Programmable Gate Arrays (FPGAs) and embedded devices. The methodology of adapting kernel shapes shows premise to be generalised in the future.

 \section*{Acknowledgement}

This project was undertaken on the Viking Cluster, which is a high performance compute facility provided by the University of York. We are grateful for computational support from the University of York High Performance Computing service, Viking and the Research Computing team.


\bibliographystyle{apalike}

\bibliography{reference}

\pagebreak


\appendices

\section{Two-Objective Optimisation}
\label{2_d}

\begin{figure*}[h!]
	\centering
	\subfloat[\label{6a}]{
		\includegraphics[width=0.48\linewidth]{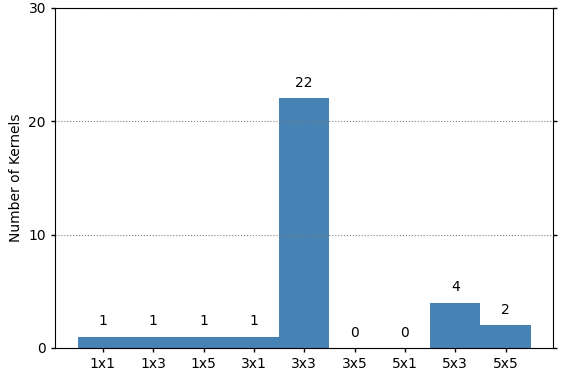}
	}
	\hfill
	\subfloat[\label{6b}]{
		\includegraphics[width=0.48\linewidth]{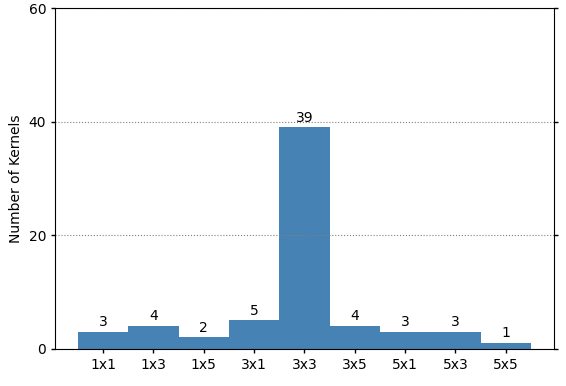}
	}
	\hfill
	\subfloat[\label{6c}]{
		\includegraphics[width=0.48\linewidth]{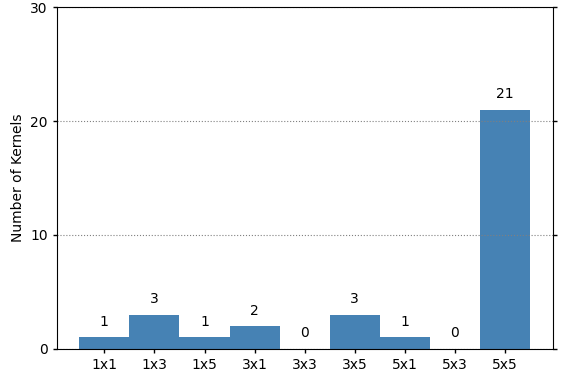}
	}
	\hfill
	\subfloat[\label{6d}]{
		\includegraphics[width=0.48\linewidth]{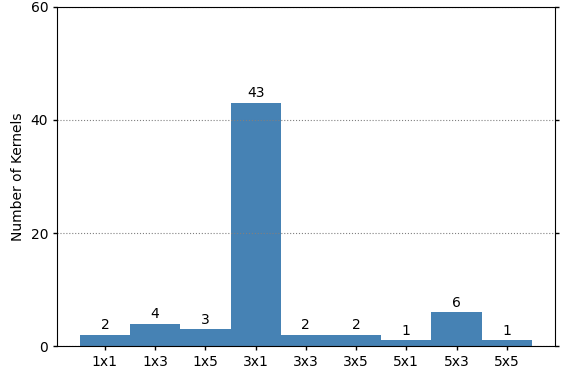}
	}
	\hfill
	\subfloat[\label{6e}]{
		\includegraphics[width=0.48\linewidth]{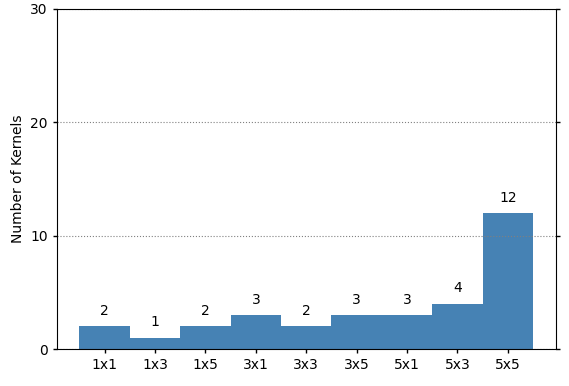}
	}
	\hfill
	\subfloat[\label{6f}]{
		\includegraphics[width=0.48\linewidth]{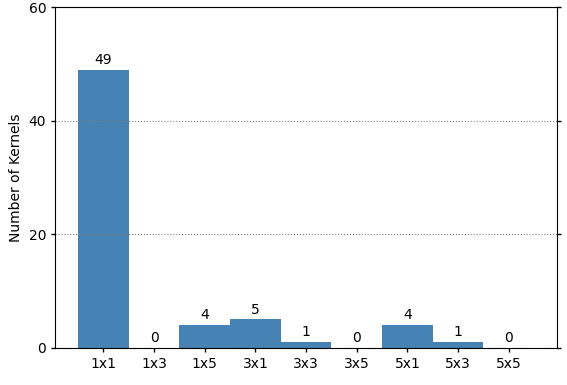}
	}
	
	\caption{(a) Kernel distribution of the first convolutional layer of Ref 1 on Fashion-MNIST dataset. (b) Kernel distribution of the second convolutional layer of Ref 1 on Fashion-MNIST dataset. (c) Kernel distribution of the first convolutional layer of Ref 2 on Fashion-MNIST dataset. (d) Kernel distribution of the second convolutional layer of Ref 2 on Fashion-MNIST dataset. (e) Kernel distribution of the first convolutional layer of Ref 3 on Fashion-MNIST dataset. (f) Kernel distribution of the second convolutional layer of Ref 3 on Fashion-MNIST dataset. }
	\label{fig_6}
\end{figure*}

These figures are for two-objective optimisation experiments in Section 4.2.
\begin {itemize}
\item Fig.~\ref{fig_1} shows the optimisation result by the proposed method and a brief plot of kernel shape distributions of three reference architectures from the optimisation result of benchmark CNN on CIFAR-10 dataset. 
\item Fig.~\ref{fig_2} shows the optimisation results by the proposed method and a brief plot of kernel shape distributions of three reference architectures from the optimisation result of benchmark CNN on MNIST dataset. 
\item Fig.~\ref{fig_3} shows the optimisation results by the proposed method and a brief plot of kernel shape distributions of three reference architectures from the optimisation results of benchmark CNN on Fashion-MNIST dataset.
\item Fig.~\ref{fig_4} gives the details of kernel usage of three reference architectures from the optimisation results of benchmark CNN on CIFAR-10 dataset. 
\item Fig.~\ref{fig_5} gives the details of kernel usage of three reference architectures from the optimisation results of benchmark CNN on MNIST dataset. 
\item Fig.~\ref{fig_6} gives the details of kernel usage of three reference architectures from the optimisation results of benchmark CNN on Fashion-MNIST dataset.
\end {itemize}

\section{Three-Objective Optimisation}
\label{3_d}
\begin{figure*}[h!]
	\centering
	\subfloat[\label{7a}]{
		\includegraphics[width=0.48\linewidth]{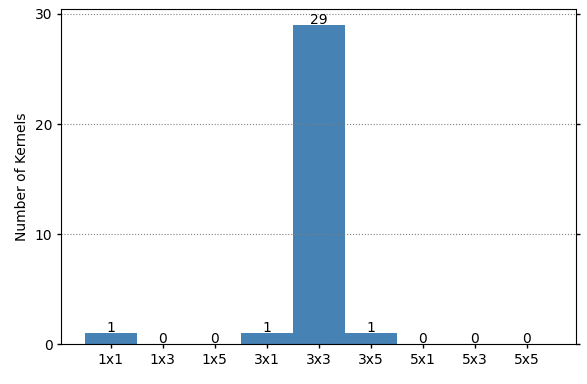}
	}
	\hfill
	\subfloat[\label{7b}]{
		\includegraphics[width=0.48\linewidth]{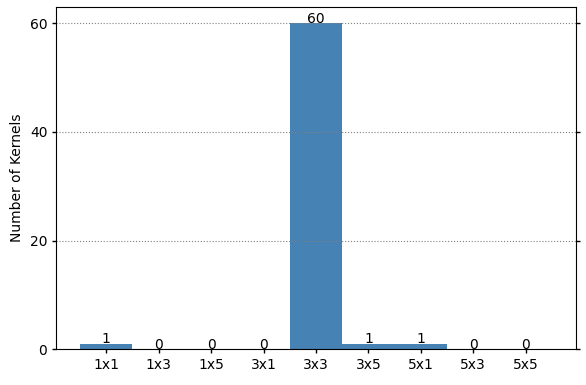}
	}
	\hfill
	\subfloat[\label{7c}]{
		\includegraphics[width=0.48\linewidth]{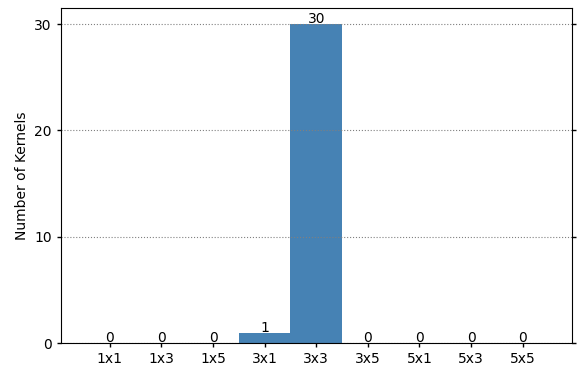}
	}
	\hfill
	\subfloat[\label{7d}]{
		\includegraphics[width=0.48\linewidth]{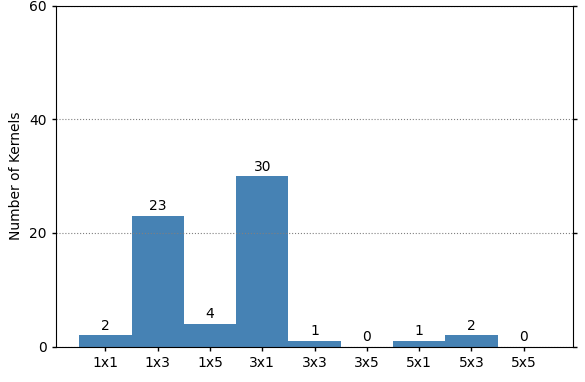}
	}
	\hfill
	\subfloat[\label{7e}]{
		\includegraphics[width=0.48\linewidth]{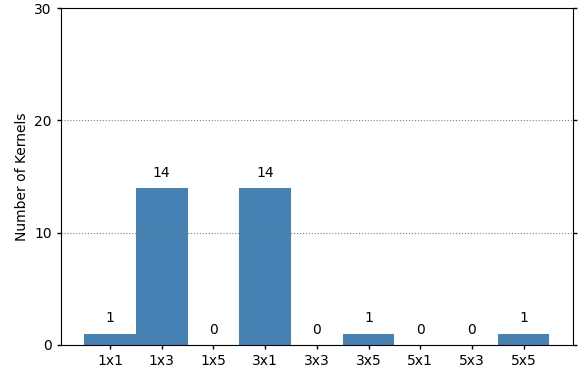}
	}
	\hfill
	\subfloat[\label{7f}]{
		\includegraphics[width=0.48\linewidth]{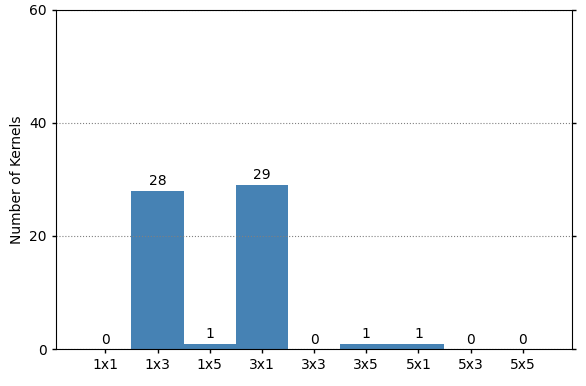}
	}
	\caption{(a) Kernel distribution of the first convolutional layer of Ref 1 on CIFAR-10 dataset. (b) Kernel distribution of the second convolutional layer of Ref 1 on CIFAR-10 dataset. (c) Kernel distribution of the first convolutional layer of Ref 2 on CIFAR-10 dataset. (d) Kernel distribution of the second convolutional layer of Ref 2 on CIFAR-10 dataset. (e) Kernel distribution of the first convolutional layer of Ref 3 on CIFAR-10 dataset. (f) Kernel distribution of the second convolutional layer of Ref 3 on CIFAR-10 dataset. }
	\label{fig_7}
\end{figure*}

These figures are for three-objective optimisation experiments in Section 4.3.
\begin {itemize}
\item Fig.~\ref{fig_7} gives the details of kernel usage of three reference architectures from the optimisation results of benchmark CNN on CIFAR-10 dataset. 
\item Fig.~\ref{fig_8} gives the details of kernel usage of three reference architectures from the optimisation results of benchmark CNN on MNIST dataset. 
\item Fig.~\ref{fig_9} gives the details of kernel usage of three reference architectures from the optimisation results of benchmark CNN on Fashion-MNIST dataset.
\end {itemize}

\section{Three-layer and Four-layer Networks Optimisation}
\label{deeper}
These figures are for three-objective optimisation of deeper CNNs with three layers. Experiments are in Section 4.4.

\begin{figure*}[h!]
	\centering
	\subfloat[\label{12a}]{
		\includegraphics[width=0.63\linewidth]{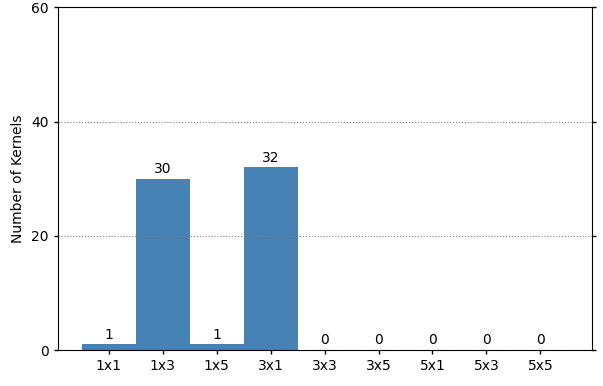}
	}
	\hfill
	\subfloat[\label{12b}]{
		\includegraphics[width=0.63\linewidth]{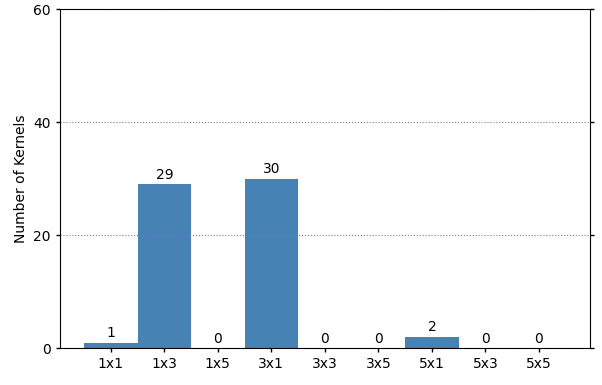}
	}
	\hfill
	\subfloat[\label{12c}]{
		\includegraphics[width=0.63\linewidth]{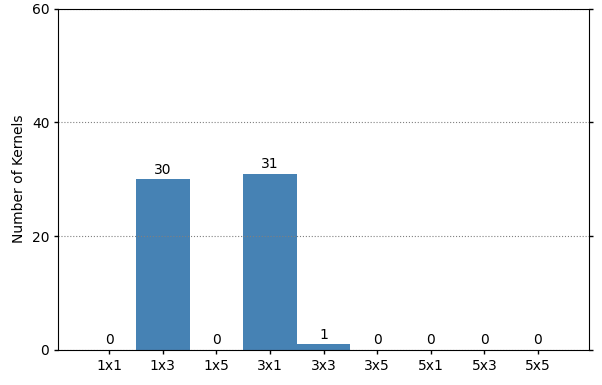}
	}
	\caption{(a) Kernel distribution of the first convolutional layer of Ref 3 from three-layer network optimisation results. (b) Kernel distribution of the second convolutional layer network of Ref 3 from three-layer optimisation results. (c) Kernel distribution of the third convolutional layer network of Ref 3 from three-layer optimisation results.
	}
	\label{fig_12}
\end{figure*}

\begin {itemize}
\item Fig.~\ref{fig_10} shows the details of kernel usage of first reference point, Ref 1, from the optimisation results of three-layer network on CIFAR-10 dataset. 
\item Fig.~\ref{fig_11} shows the details of kernel usage of second reference point, Ref 2, from the optimisation results of three-layer network on CIFAR-10 dataset.  
\item Fig.~\ref{fig_12} shows the details of kernel usage of third reference point, Ref 3, from the optimisation results of three-layer network on CIFAR-10 dataset. 
\end {itemize}

These figures are for three-objective optimisation of deeper CNNs with four layers. Experiments are in Section 4.4.

\begin{figure*}[h!]
	\centering
	\subfloat[\label{13a}]{
		\includegraphics[width=0.485\linewidth]{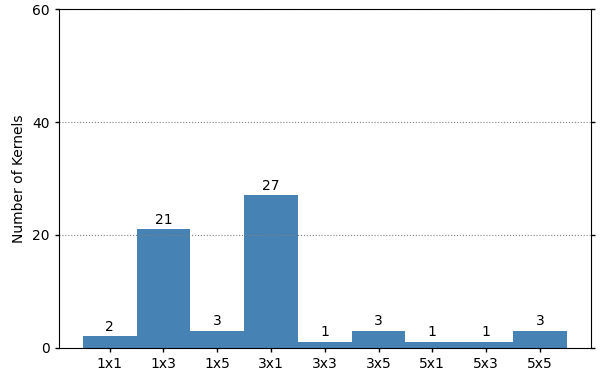}
	}
	\hfill
	\subfloat[\label{13b}]{
		\includegraphics[width=0.485\linewidth]{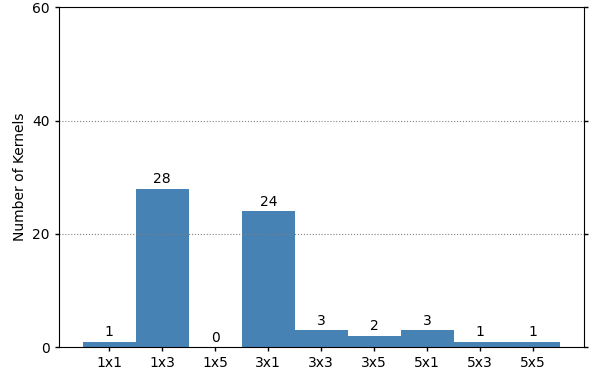}
	}
	
	\subfloat[\label{13c}]{
		\includegraphics[width=0.485\linewidth]{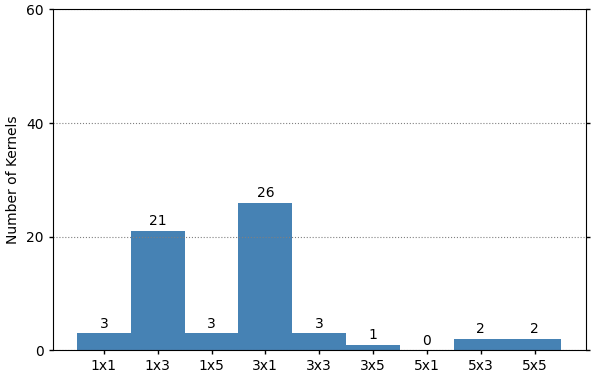}
	}
	\hfill
	\subfloat[\label{13d}]{
		\includegraphics[width=0.485\linewidth]{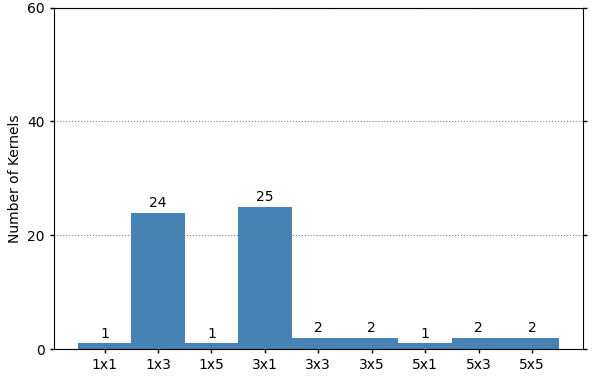}
	}
	\caption{(a) Kernel distribution of the first convolutional layer of Ref 1 from four-layer network optimisation results. (b) Kernel distribution of the second convolutional layer of Ref 1 from four-layer network optimisation results. (c) Kernel distribution of the third convolutional layer of Ref 1 from four-layer network optimisation results. (d) Kernel distribution of the fourth convolutional layer network of Ref 1 from three-layer optimisation results.
	}
	\label{fig_13}
\end{figure*}

\begin{figure*}[h!]
	\centering
	\subfloat[\label{14a}]{
		\includegraphics[width=0.485\linewidth]{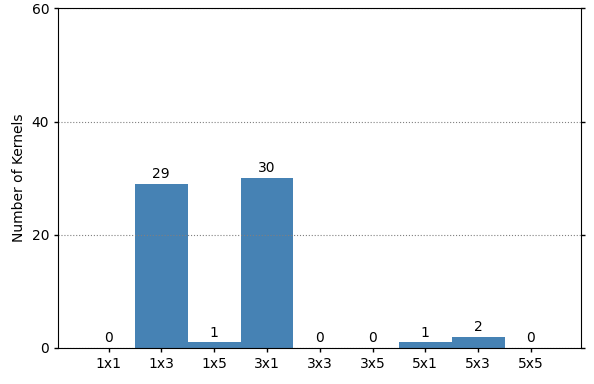}
	}
	\hfill
	\subfloat[\label{14b}]{
		\includegraphics[width=0.485\linewidth]{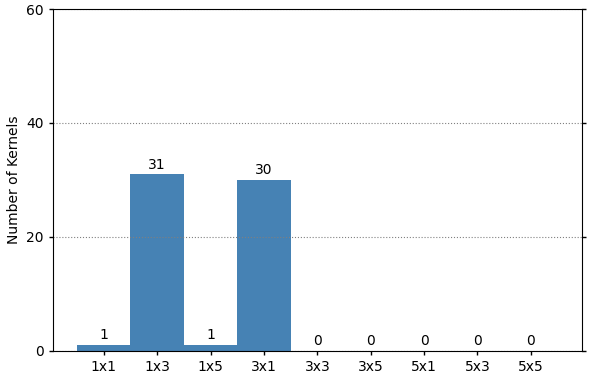}
	}
	
	\subfloat[\label{14c}]{
		\includegraphics[width=0.485\linewidth]{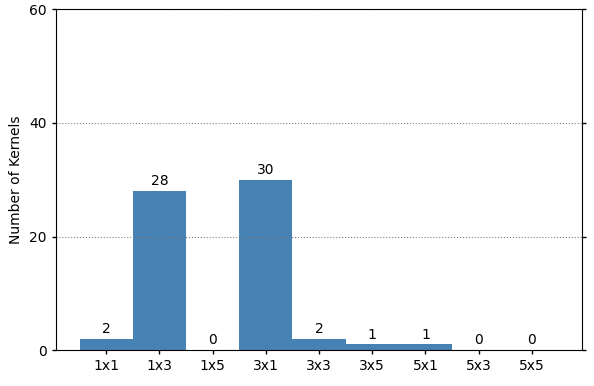}
	}
	\hfill
	\subfloat[\label{14d}]{
		\includegraphics[width=0.48\linewidth]{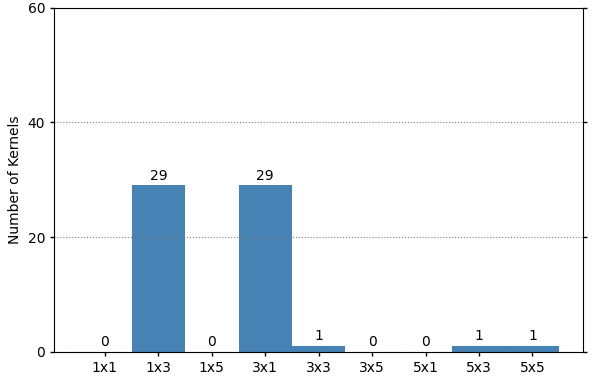}
	}
	\caption{(a) Kernel distribution of the first convolutional layer of Ref 2 from four-layer network optimisation results. (b) Kernel distribution of the second convolutional layer of Ref 2 from four-layer network optimisation results. (c) Kernel distribution of the third convolutional layer of Ref 2 from four-layer network optimisation results. (d) Kernel distribution of the fourth convolutional layer network of Ref 2 from three-layer optimisation results.
	}
	\label{fig_14}
\end{figure*}

\begin{figure*}[h!]
	\centering
	\subfloat[\label{15a}]{
		\includegraphics[width=0.485\linewidth]{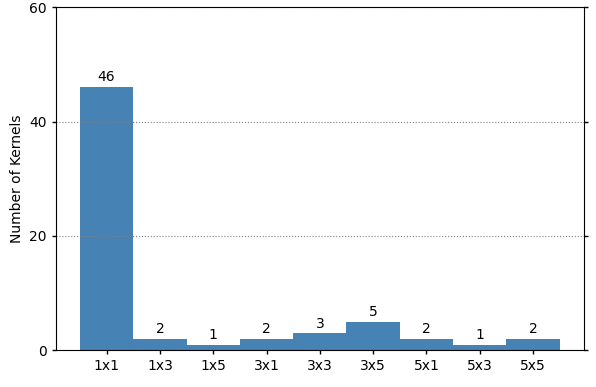}
	}
	\hfill
	\subfloat[\label{15b}]{
		\includegraphics[width=0.485\linewidth]{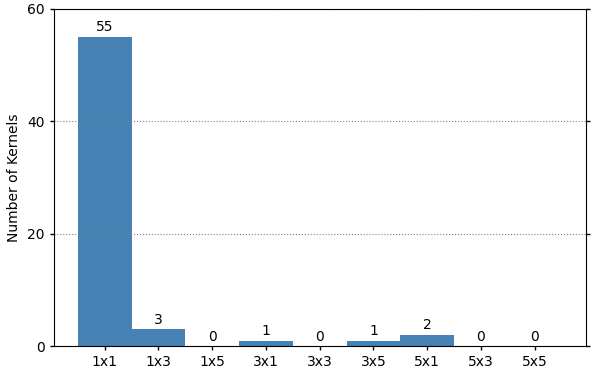}
	}
	
	\subfloat[\label{15c}]{
		\includegraphics[width=0.485\linewidth]{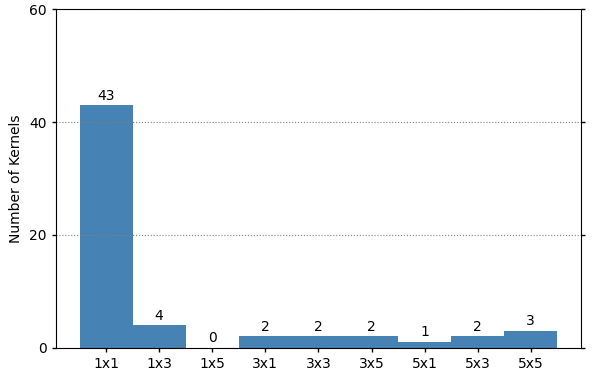}
	}
	\hfill
	\subfloat[\label{15d}]{
		\includegraphics[width=0.485\linewidth]{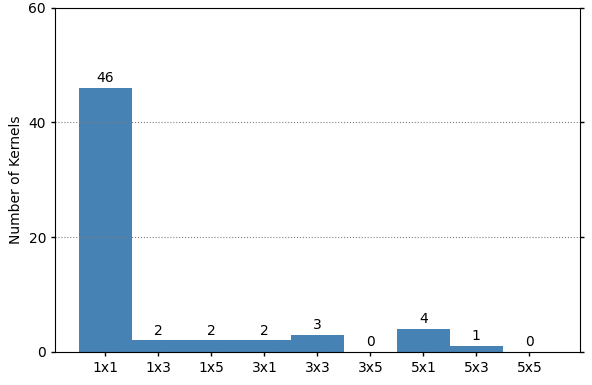}
	}
	\caption{(a) Kernel distribution of the first convolutional layer of Ref 3 from four-layer network optimisation results. (b) Kernel distribution of the second convolutional layer of Ref 3 from four-layer network optimisation results. (c) Kernel distribution of the third convolutional layer of Ref 3 from four-layer network optimisation results. (d) Kernel distribution of the fourth convolutional layer network of Ref 3 from three-layer optimisation results.
	}
	\label{fig_15}
\end{figure*}

\begin {itemize}
\item Fig.~\ref{fig_13} shows the details of kernel usage of first reference point, Ref 1, from the optimisation results of four-layer network on CIFAR-10 dataset. 
\item Fig.~\ref{fig_14} shows the details of kernel usage of second reference point, Ref 2, from the optimisation results of four-layer network on CIFAR-10 dataset.  
\item Fig.~\ref{fig_15} shows the details of kernel usage of third reference point, Ref 3, from the optimisation results of four-layer network on CIFAR-10 dataset. 
\end {itemize}
\clearpage


\begin{figure*}[h!]
	\centering
	\subfloat[\label{1a}]{
		\includegraphics[width=0.65\linewidth]{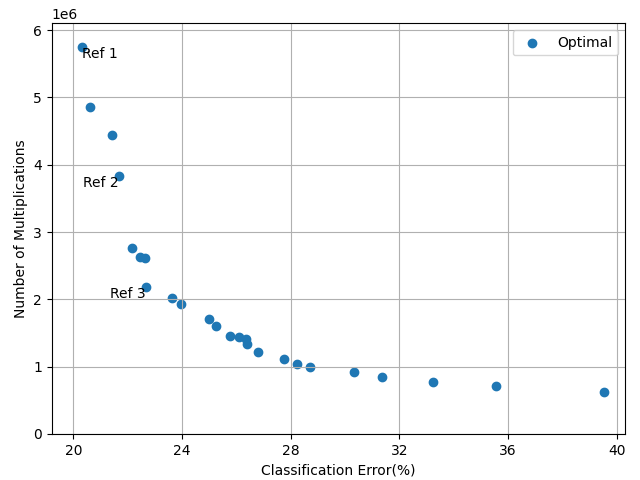}
	}
	\hfill
	\subfloat[\label{1b}]{
		\includegraphics[width=0.8\linewidth]{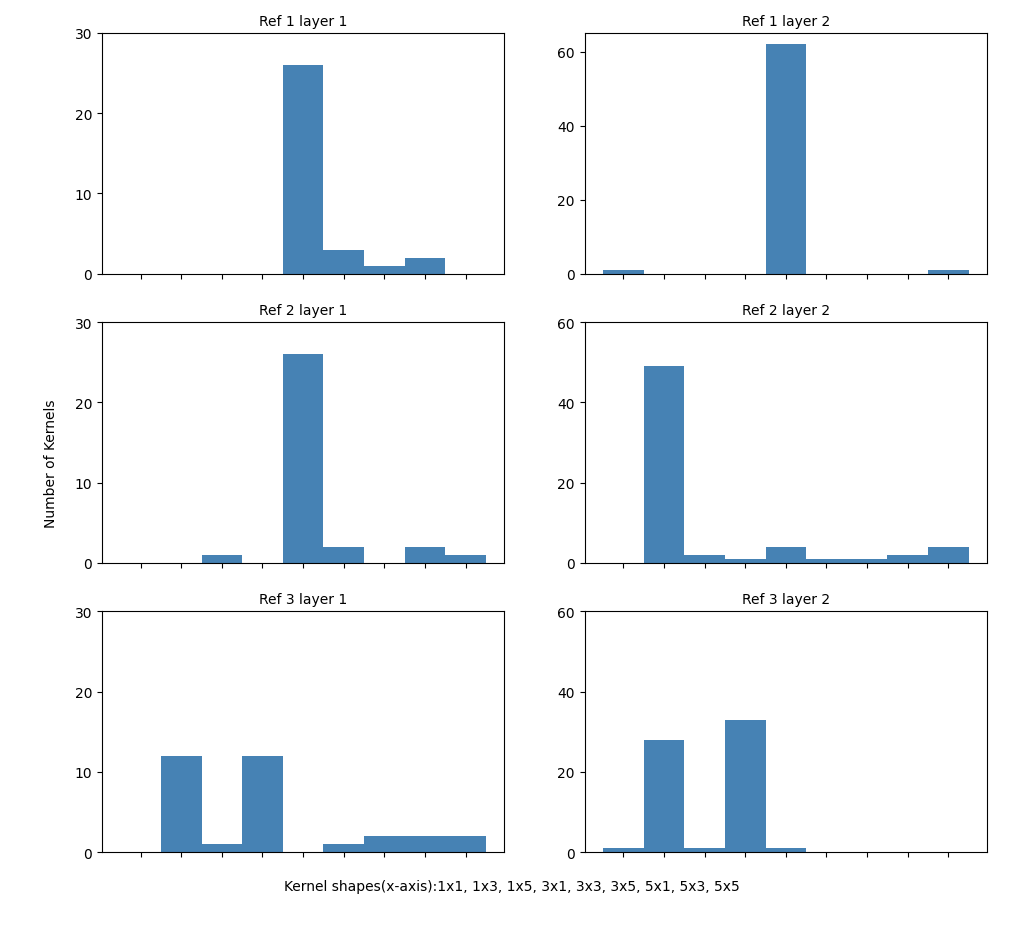}
	}
	\caption{(a) Optimised architectures by the proposed method after 100 generations on CIFAR-10. (b) The kernel distribution of three reference points on (a). }
	\label{fig_1}
\end{figure*}


\begin{figure*}[h!]
	\centering
	\subfloat[\label{2a}]{
		\includegraphics[width=0.65\linewidth]{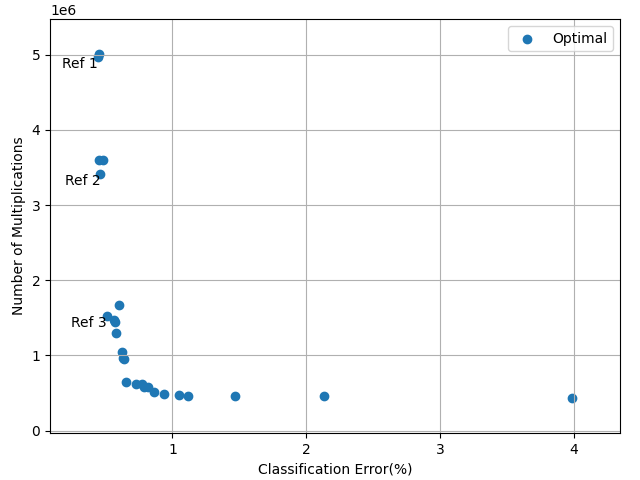}
	}
	\hfill
	\subfloat[\label{2b}]{
		\includegraphics[width=0.85\linewidth]{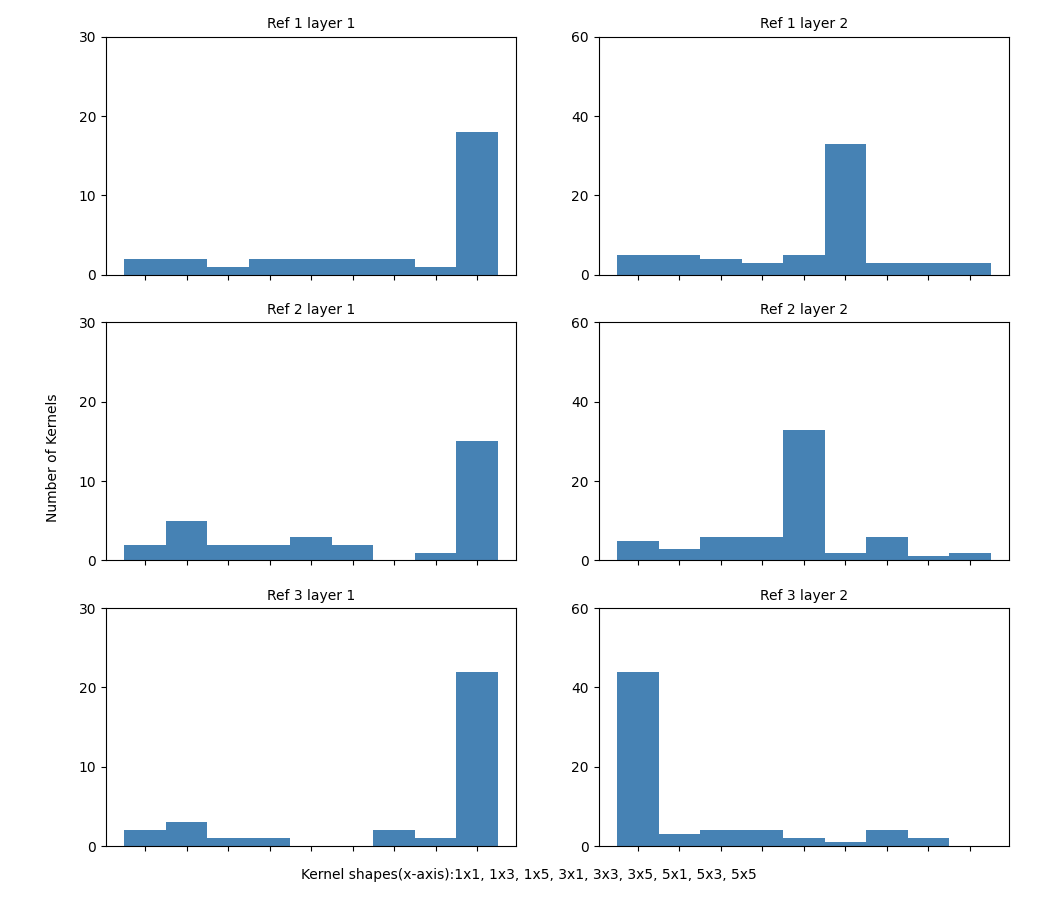}
	}
	\caption{(a) Optimised architectures by the proposed method after 100 generations on MNIST. (b) The kernel distribution of three reference points on (a). }
	\label{fig_2}
\end{figure*}


\begin{figure*}[h!]
	\centering
	\subfloat[\label{3a}]{
		\includegraphics[width=0.65\linewidth]{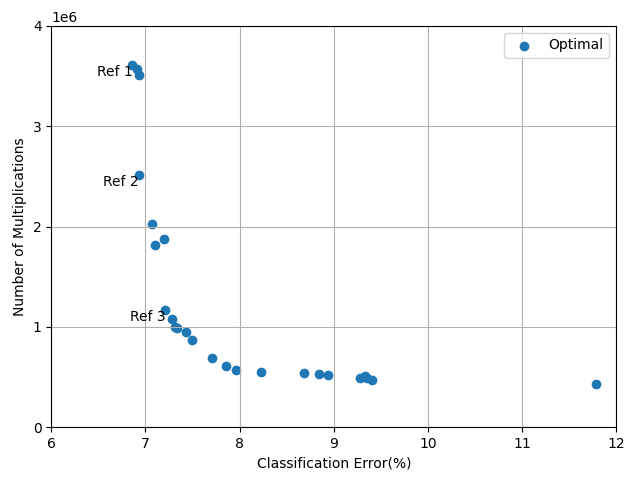}
	}
	\hfill
	\subfloat[\label{3b}]{
		\includegraphics[width=0.85\linewidth]{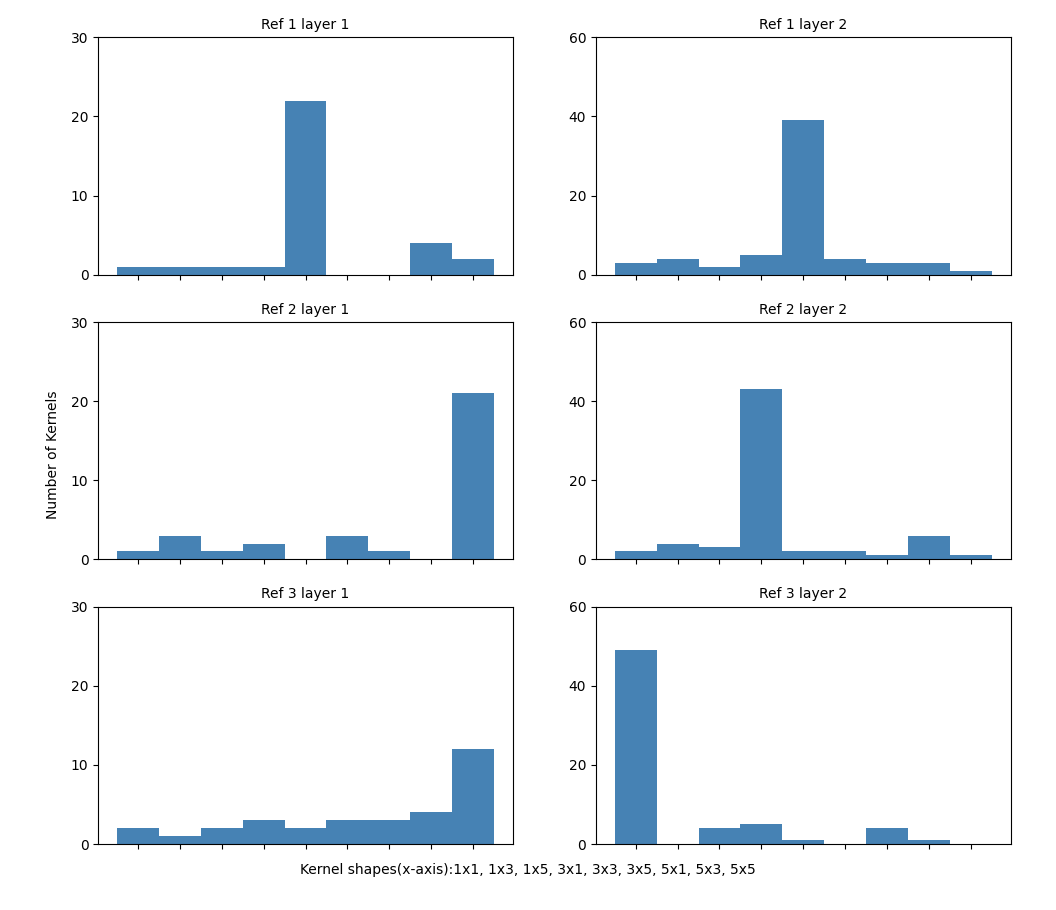}
	}
	\caption{(a) Optimised architectures by the proposed method after 100 generations on Fashion-MNIST. (b) The kernel distribution of three reference points on (a). }
	\label{fig_3}
\end{figure*}

\begin{figure*}[h!]
	\centering
	\subfloat[\label{4a}]{
		\includegraphics[width=0.48\linewidth]{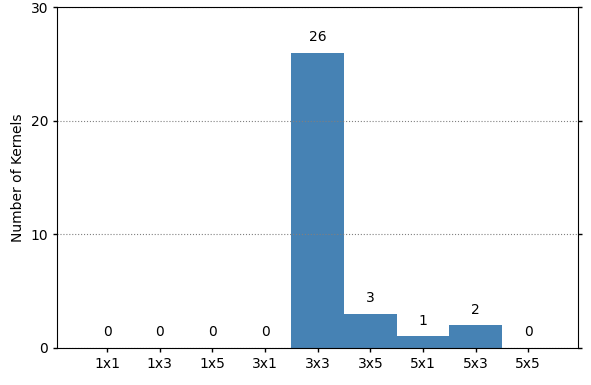}
	}
	\hfill
	\subfloat[\label{4b}]{
		\includegraphics[width=0.48\linewidth]{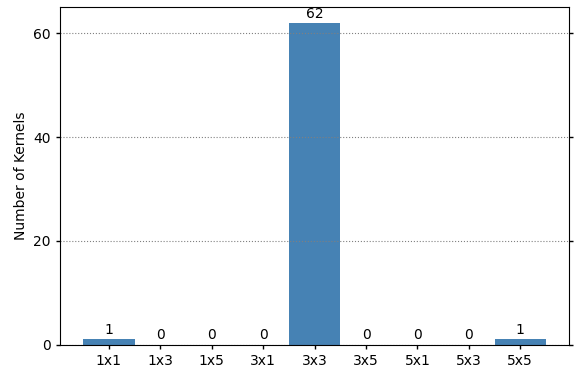}
	}
	\hfill
	\subfloat[\label{4c}]{
		\includegraphics[width=0.48\linewidth]{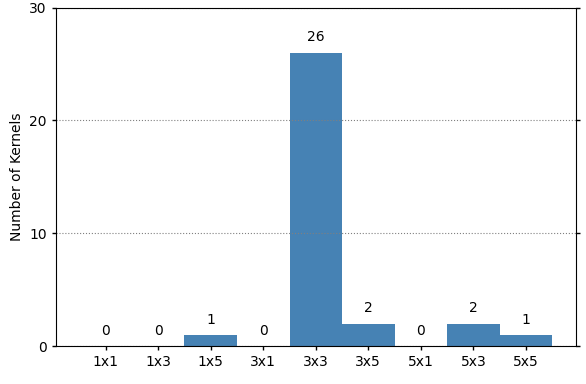}
	}
	\hfill
	\subfloat[\label{4d}]{
		\includegraphics[width=0.48\linewidth]{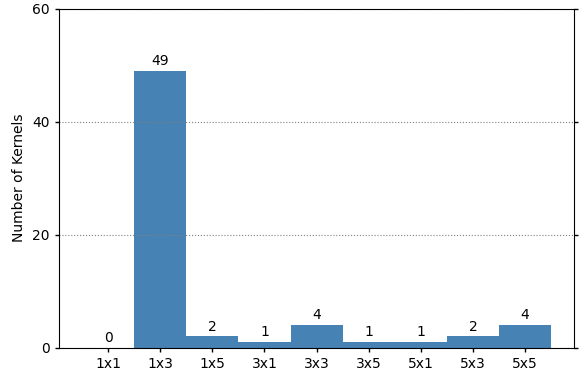}
	}
	\hfill
	\subfloat[\label{4e}]{
		\includegraphics[width=0.48\linewidth]{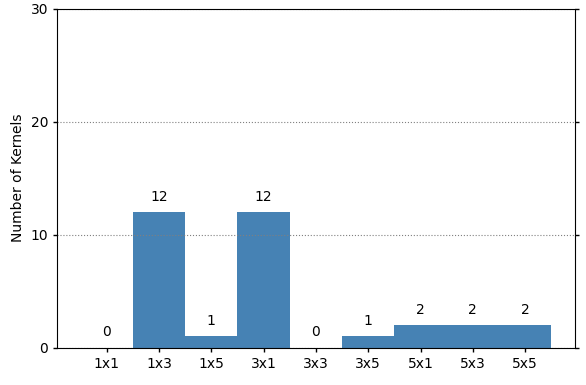}
	}
	\hfill
	\subfloat[\label{4f}]{
		\includegraphics[width=0.48\linewidth]{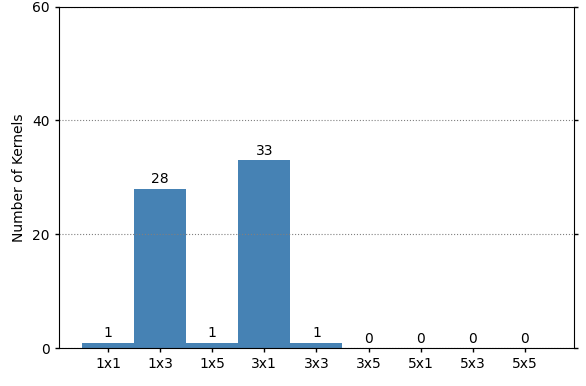}
	}
	
	\caption{(a) Kernel distribution of the first convolutional layer of Ref 1 on CIFAR-10 dataset. (b) Kernel distribution of the second convolutional layer of Ref 1 on CIFAR-10 dataset. (c) Kernel distribution of the first convolutional layer of Ref 2 on CIFAR-10 dataset. (d) Kernel distribution of the second convolutional layer of Ref 2 on CIFAR-10 dataset. (e) Kernel distribution of the first convolutional layer of Ref 3 on CIFAR-10 dataset. (f) Kernel distribution of the second convolutional layer of Ref 3 on CIFAR-10 dataset. }
	\label{fig_4}
\end{figure*}

\begin{figure*}[h!]
	\centering
	\subfloat[\label{5a}]{
		\includegraphics[width=0.48\linewidth]{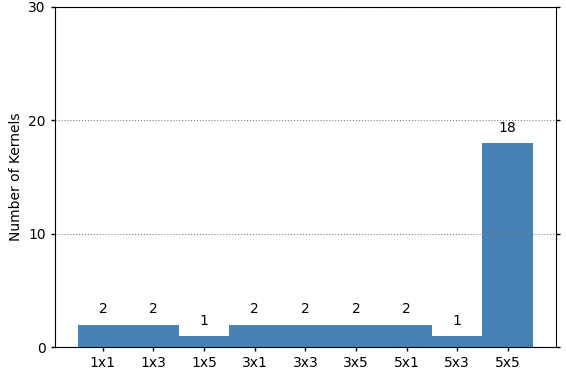}
	}
	\hfill
	\subfloat[\label{5b}]{
		\includegraphics[width=0.48\linewidth]{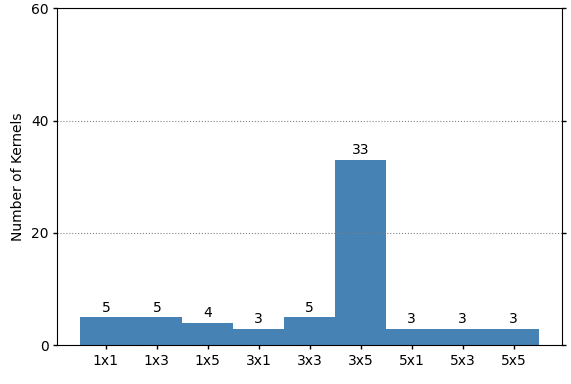}
	}
	\hfill
	\subfloat[\label{5c}]{
		\includegraphics[width=0.48\linewidth]{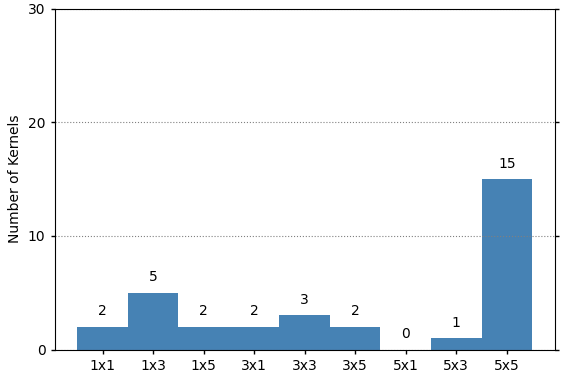}
	}
	\hfill
	\subfloat[\label{5d}]{
		\includegraphics[width=0.48\linewidth]{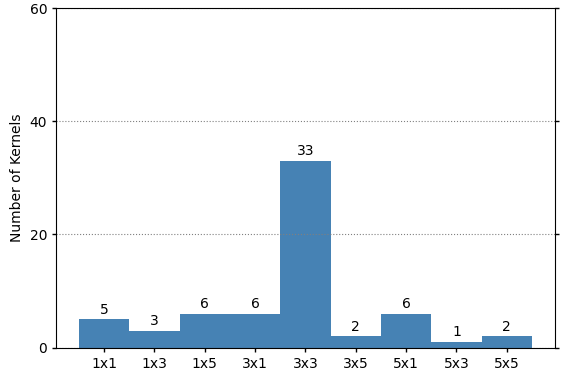}
	}
	\hfill
	\subfloat[\label{5e}]{
		\includegraphics[width=0.48\linewidth]{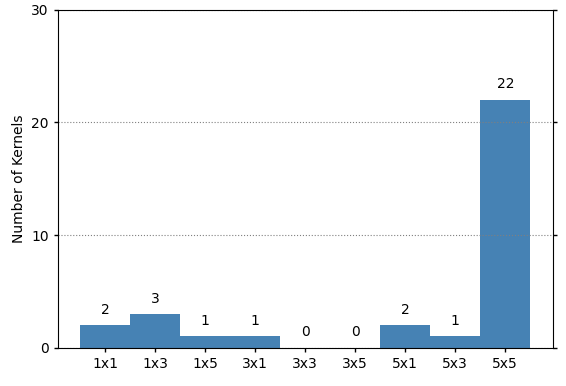}
	}
	\hfill
	\subfloat[\label{5f}]{
		\includegraphics[width=0.48\linewidth]{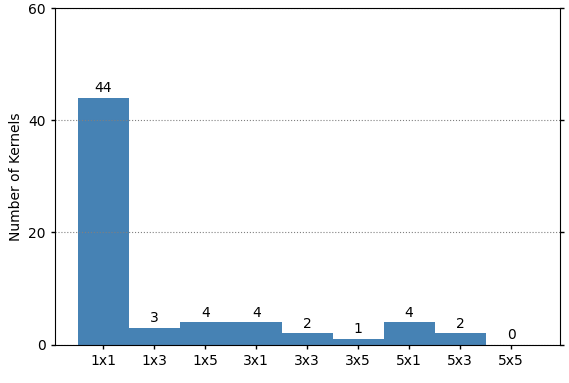}
	}
	
	\caption{(a) Kernel distribution of the first convolutional layer of Ref 1 on MNIST dataset. (b) Kernel distribution of the second convolutional layer of Ref 1 on MNIST dataset. (c) Kernel distribution of the first convolutional layer of Ref 2 on MNIST dataset. (d) Kernel distribution of the second convolutional layer of Ref 2 on MNIST dataset. (e) Kernel distribution of the first convolutional layer of Ref 3 on MNIST dataset. (f) Kernel distribution of the second convolutional layer of Ref 3 on MNIST dataset. }
	\label{fig_5}
\end{figure*}

\begin{figure*}[h!]
	\centering
	\subfloat[\label{8a}]{
		\includegraphics[width=0.48\linewidth]{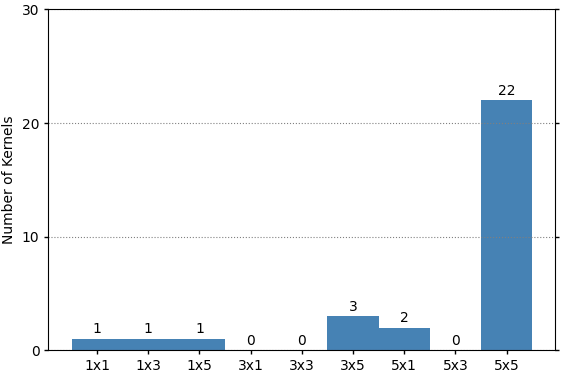}
	}
	\hfill
	\subfloat[\label{8b}]{
		\includegraphics[width=0.48\linewidth]{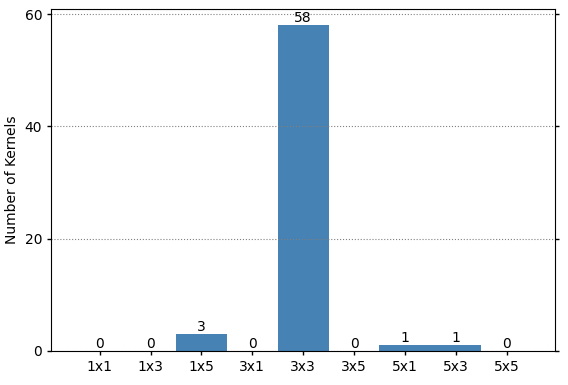}
	}
	\hfill
	\subfloat[\label{8c}]{
		\includegraphics[width=0.48\linewidth]{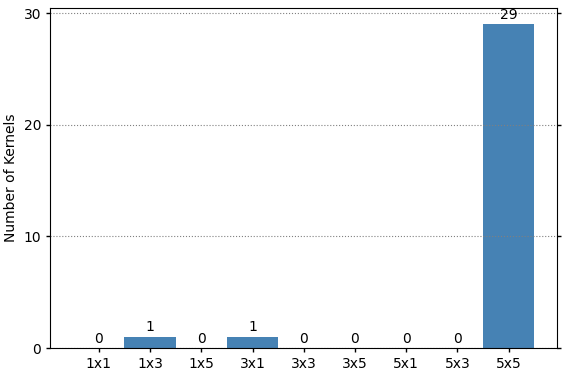}
	}
	\hfill
	\subfloat[\label{8d}]{
		\includegraphics[width=0.48\linewidth]{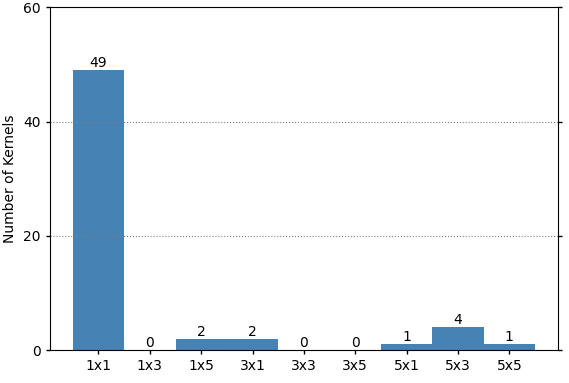}
	}
	\hfill
	\subfloat[\label{8e}]{
		\includegraphics[width=0.48\linewidth]{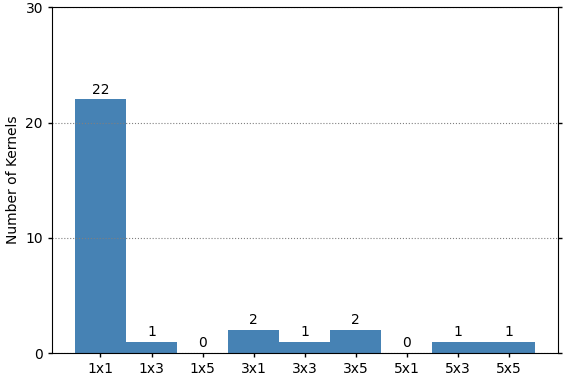}
	}
	\hfill
	\subfloat[\label{8f}]{
		\includegraphics[width=0.48\linewidth]{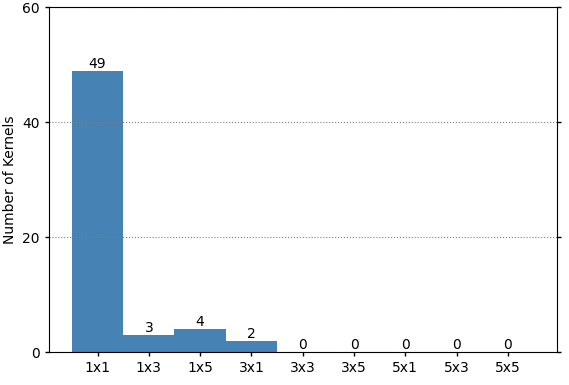}
	}
	\caption{(a) Kernel distribution of the first convolutional layer of Ref 1 on MNIST dataset. (b) Kernel distribution of the second convolutional layer of Ref 1 on MNIST dataset. (c) Kernel distribution of the first convolutional layer of Ref 2 on MNIST dataset. (d) Kernel distribution of the second convolutional layer of Ref 2 on MNIST dataset. (e) Kernel distribution of the first convolutional layer of Ref 3 on MNIST dataset. (f) Kernel distribution of the second convolutional layer of Ref 3 on MNIST dataset. }
	\label{fig_8}
\end{figure*}

\begin{figure*}[h!]
	\centering
	\subfloat[\label{9a}]{
		\includegraphics[width=0.48\linewidth]{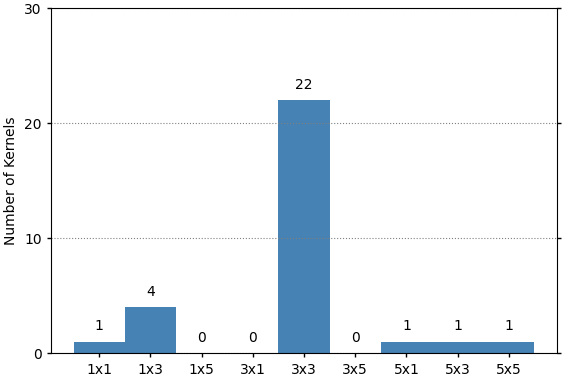}
	}
	\hfill
	\subfloat[\label{9b}]{
		\includegraphics[width=0.48\linewidth]{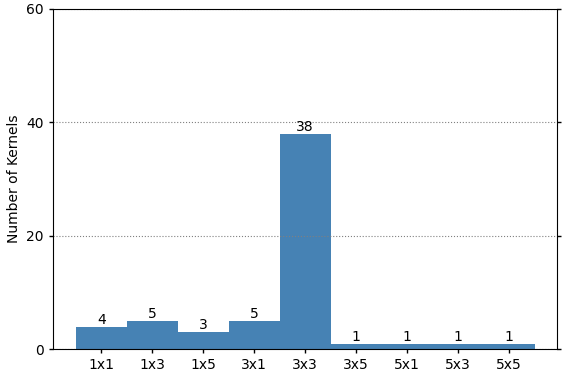}
	}
	\hfill
	\subfloat[\label{9c}]{
		\includegraphics[width=0.48\linewidth]{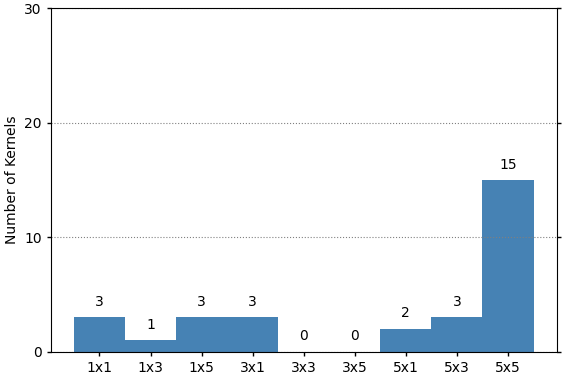}
	}
	\hfill
	\subfloat[\label{9d}]{
		\includegraphics[width=0.48\linewidth]{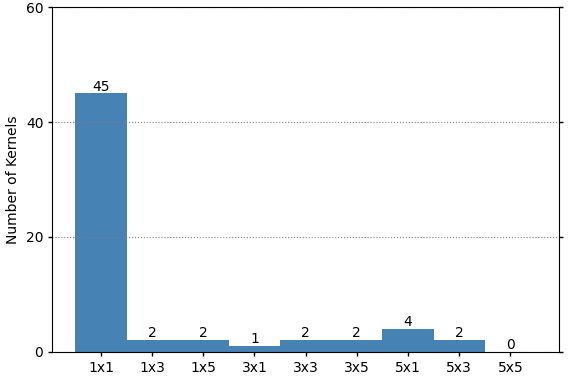}
	}
	\hfill
	\subfloat[\label{9e}]{
		\includegraphics[width=0.48\linewidth]{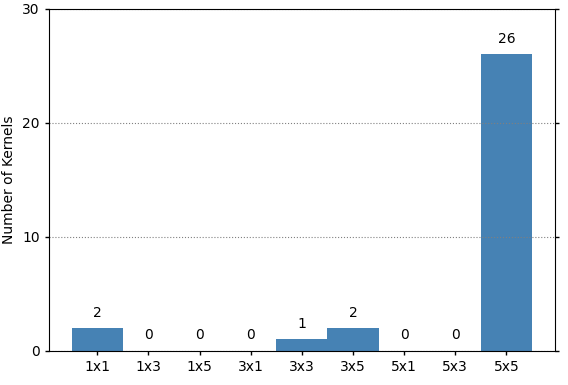}
	}
	\hfill
	\subfloat[\label{9f}]{
		\includegraphics[width=0.48\linewidth]{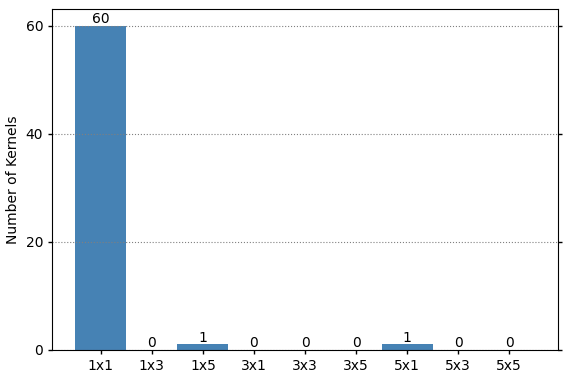}
	}
	\caption{(a) Kernel distribution of the first convolutional layer of Ref 1 on Fashion-MNIST dataset. (b) Kernel distribution of the second convolutional layer of Ref 1 on Fashion-MNIST dataset. (c) Kernel distribution of the first convolutional layer of Ref 2 on Fashion-MNIST dataset. (d) Kernel distribution of the second convolutional layer of Ref 2 on Fashion-MNIST dataset. (e) Kernel distribution of the first convolutional layer of Ref 3 on Fashion-MNIST dataset. (f) Kernel distribution of the second convolutional layer of Ref 3 on Fashion-MNIST dataset. }
	\label{fig_9}
\end{figure*}
\begin{figure*}[h!]
	\centering
	\subfloat[\label{10a}]{
		\includegraphics[width=0.63\linewidth]{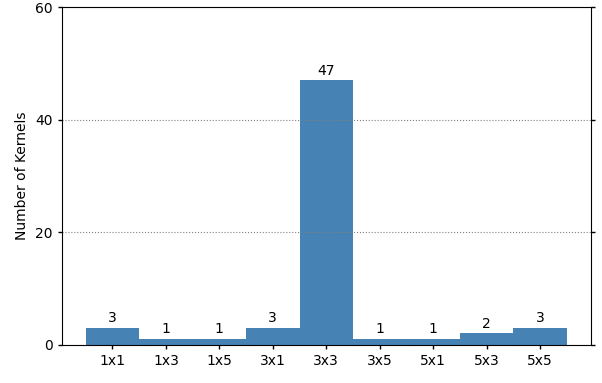}
	}
	\hfill
	\subfloat[\label{10b}]{
		\includegraphics[width=0.63\linewidth]{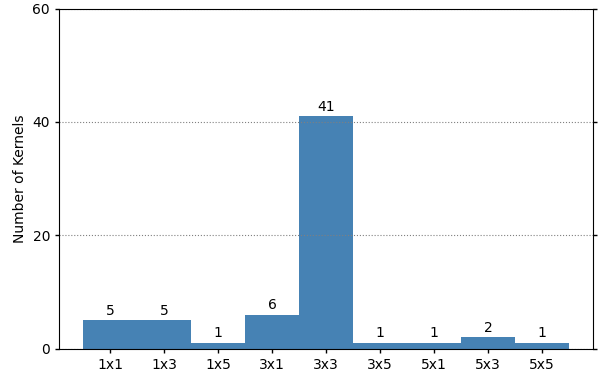}
	}
	\hfill
	\subfloat[\label{10c}]{
		\includegraphics[width=0.63\linewidth]{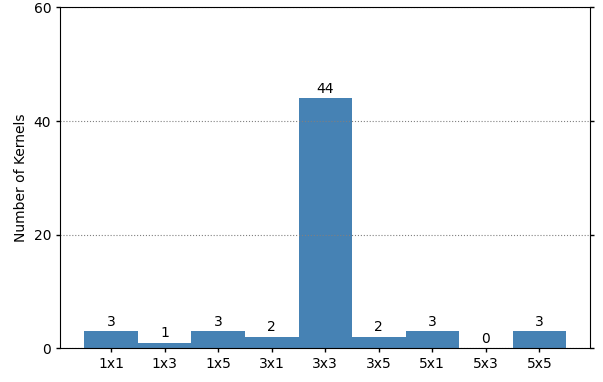}
	}
	\caption{(a) Kernel distribution of the first convolutional layer of Ref 1 from three-layer network optimisation results. (b) Kernel distribution of the second convolutional layer of Ref 1 from three-layer network optimisation results. (c) Kernel distribution of the third convolutional layer of Ref 1 from three-layer network optimisation results.
	}
	\label{fig_10}
\end{figure*}

\begin{figure*}[h!]
	\centering
	\subfloat[\label{11a}]{
		\includegraphics[width=0.63\linewidth]{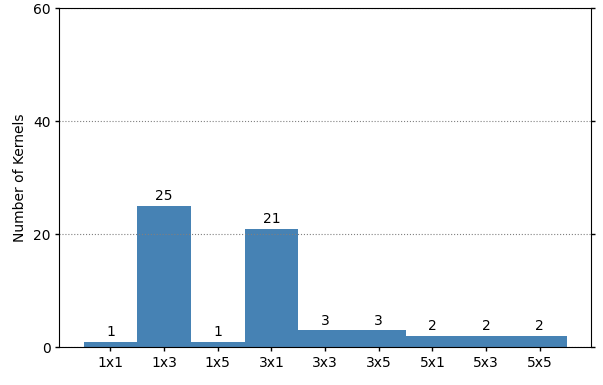}
	}
	\hfill
	\subfloat[\label{11b}]{
		\includegraphics[width=0.63\linewidth]{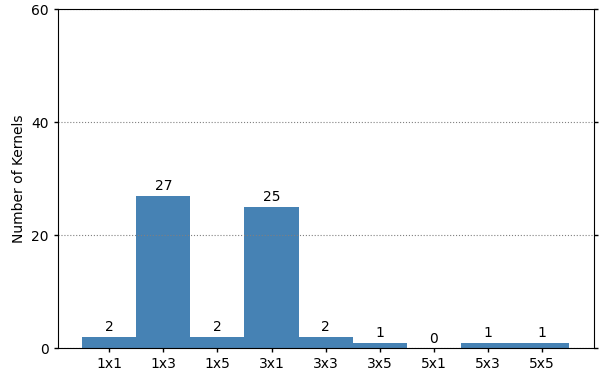}
	}
	\hfill
	\subfloat[\label{11c}]{
		\includegraphics[width=0.63\linewidth]{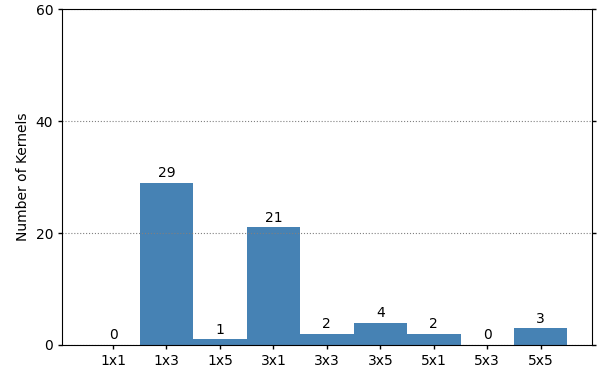}
	}
	\caption{(a) Kernel distribution of the first convolutional layer of Ref 2 from three-layer network optimisation results. (b) Kernel distribution of the second convolutional layer of Ref 2 from three-layer network optimisation results. (c) Kernel distribution of the third convolutional layer of Ref 2 from three-layer network optimisation results.
	}
	\label{fig_11}
\end{figure*}

\end{document}